# Mixed-Effects Location Scale Models for Joint Modelling School Value-Added Effects on the Mean and Variance of Student Achievement

George Leckie, Richard Parker, Harvey Goldstein, Kate Tilling

Centre for Multilevel Modelling

School of Education

University of Bristol, UK

**Corresponding author**

Professor George Leckie

Centre for Multilevel Modelling

School of Education

University of Bristol

35 Berkeley Square

Bristol

BS8 1JA

United Kingdom

g.leckie@bristol.ac.uk

**Authors**

GEORGE LECKIE is a Professor of Social Statistics at the Centre for Multilevel Modelling and School of Education, University of Bristol, Bristol, United Kingdom; e-mail: g.leckie@bristol.ac.uk. His research interests are in the development, application, and dissemination of multilevel models to analyze messy and complex clustered cross-sectional and longitudinal data in educational and other social science research.

RICHARD PARKER is a Senior Research Associate in Applied Statistics / Epidemiology at the Bristol Medical School, University of Bristol, Bristol, United Kingdom; e-mail: richard.parker@bristol.ac.uk. His research interests are in multilevel and mixed-effects location scale models and their applications in epidemiology.

HARVEY GOLDSTEIN died 9th April 2020 aged 80 while this article was in preparation. He was a Professor of Social Statistics at the Centre for Multilevel Modelling and School of Education, University of Bristol, Bristol, United Kingdom. His research interests were in the use of statistical modelling techniques in the construction and analysis of educational tests, educational (school) effectiveness, the methodology of multilevel modelling, and Bayesian modelling methods for handling missing data values and measurement errors.

KATE TILLING is a Professor of Medical Statistics at the Bristol Medical School, University of Bristol, United Kingdom; e-mail: kate.tilling@bristol.ac.uk. Her research interests are in the development and application of statistical methods to causal problems in epidemiology/health services research. Two particular areas are methods for analysis of longitudinal data, and methods for minimizing bias due to missing data.

**Acknowledgements of support**

This research was funded by UK Economic and Social Research Council (ESRC) grants ES/R010285/1 and ES/W000555/1 and UK Medical Research Council (MRC) grant MR/N027485/1. This work contains statistical data from Office for National Statistics (ONS), UK, which is Crown Copyright. The use of the ONS statistical data in this work does not imply the endorsement of the ONS in relation to the interpretation or analysis of the statistical data.

# Mixed-Effects Location Scale Models for Joint Modeling School Value-Added Effects on the Mean and Variance of Student Achievement

## Abstract

School value-added models are widely applied to study, monitor, and hold schools to account for school differences in student learning. The traditional model is a mixed-effects linear regression of student current achievement on student prior achievement, background characteristics, and a school random intercept effect. The latter is referred to as the school value-added score and measures the mean student covariate-adjusted achievement in each school. In this article, we argue that further insights may be gained by additionally studying the variance in this quantity in each school. These include the ability to identify both individual schools and school types that exhibit unusually high or low variability in student achievement, even after accounting for differences in student intakes. We explore and illustrate how this can be done via fitting mixed-effects location scale versions of the traditional school value-added model. We discuss the implications of our work for research and school accountability systems.

## 1. Introduction

School value-added models attempt to estimate school differences in student achievement and are widely applied in educational (Goldstein, 1997; Reynolds et al., 2014; Teddlie & Reynolds, 2000) and statistical research (American Statistical Association, 2014; Braun & Wainer, 2007; McCaffrey et al., 2004; Raudenbush & Willms, 1995; Wainer, 2004). They are also used in the US, UK, and other school accountability systems where the predicted school differences, often referred to as school value-added scores, provide the basis of reward and sanction decisions on schools (Amrein-Beardsley, 2014; Castellano & Ho, 2013; Koretz, 2017; Leckie & Goldstein, 2017; OECD, 2008). In educational and statistical research, there is an additional interest in identifying school policies and practices that predict the school differences and that might therefore prove effective at raising student achievement in schools in general.

The traditional school value-added model is formulated as a mixed-effects (multilevel or hierarchical) linear regression model (Goldstein, 2011; Raudenbush & Bryk, 2002; Snijders & Bosker, 2012) of student current achievement on student prior achievement measured at the start of the value-added period (typically defined as one or more school years or a phase of schooling) and a school random intercept effect to predict the school differences (Aitkin & Longford, 1986; Goldstein et al., 1993; Raudenbush & Bryk, 1986). The adjustment for student prior achievement is fundamental as simpler comparisons of unadjusted school mean achievement would in large part reflect school differences in student achievement present at the start of the value-added period. Such differences are argued beyond the control of the school. Student sociodemographic characteristics are often added to adjust for initial school differences in student composition more convincingly (Ballou et al., 2004; Leckie & Goldstein, 2019; Leckie & Prior, 2022; Levy et al., 2023). Schools with higher scores are described as adding more value: producing higher student achievement for any given set of students. The scores are argued to reflect the net influences of differences in the quality of teaching, availability of resources, and other policies and practices across schools which are typically unobserved to the data analyst. The regression coefficient on student prior achievement is occasionally allowed to vary across schools. The resulting random slope model is sometimes referred to as a 'differential school effectiveness' model as this extension allows schools to now have different effects for different types of students (Nuttal et al., 1989; Strand, 2010; Scherer & Nilsen, 2019).

While the traditional school value-added model is widely applied (Levy et al., 2019), it is important to realize that this model is just a regression model fitted to observational data and so the effects attributed to schools may also be caused by other factors that are not captured by the model (American Statistical Association, 2014). That is, while there is consensus that the predicted school effects are fairer and more meaningful measures to compare schools than comparing simple school mean achievement scores, the additional assumptions required to interpret these predicted school effects as causal effects rather than as merely adjusted school mean differences are challenging (Amrein-Beardsley, 2019; Reardon & Raudenbush, 2009; Rubin et al., 2004). For example, the school-level exogeneity assumption (independence of covariates and school random effect) will fail if higher prior achieving students select into more effective schools, perhaps because such students are from more affluent families who are more able to buy into the catchment areas of these schools (Angrist et al., 2021; De Fraine, 2002; Thomas & Mortimore, 1996; Timmermans & Thomas, 2015). The parameter estimates of the school value-added models presented in this article should therefore be viewed as measures of association and the predicted school effects as descriptive differences in means and variances of student achievement across schools where inevitably only partial and imperfect adjustments have been made for school differences in student characteristics at intake.

In the traditional school value-added model, the difference between observed and predicted student current achievement defines the total residual, which can be viewed as a covariate-adjusted (residualized) measure of student current achievement (i.e., a controlled comparison of student achievement levels). The total residual is modelled as the summation of the school random intercept effect and the student residual. The school random effect measures the mean student adjusted achievement in each school. In contrast, the constant residual variance implicitly assumes the variance in student adjusted achievement is the same in every school. This inconsistent modeling of the mean and variance does not seem realistic. Any given school policy or practice will have different effects on students as a function of their observed and unobserved characteristics and will therefore contribute to the variance in student adjusted achievement operating in each school. Indeed, this is the motivation for the random slope extension to the traditional value-added model described above. In practice, however, this extension can only be used to account for a limited number of observed student characteristics, not to all observed and unobserved student characteristics (Raudenbush and Bryk, 2002). Thus, the different sets of school policies and practices operating in each school will lead the variance in student adjusted achievement to vary across schools, even in random slope models.

Studying the variance in student adjusted achievement in each school may therefore provide valuable new insights into the differences in student learning between schools. Consider two schools which show similarly high levels of mean student adjusted achievement. The traditional school value-added model would describe these two schools as equally effective. Suppose, however, the first school shows higher variance in their student adjusted achievement scores than the second school. Which school should now be viewed more positively? The school with the higher variance will have more students making exceptionally high adjusted achievement (a positive) albeit at the expense of more students also making unacceptably low adjusted achievement (a negative). All else equal, the school with the higher variance will also show a weaker link between prior and current achievement and so in this school low prior achievement students are more able to raise up the achievement distribution (a positive), but equally and necessarily, high prior achievement students are more likely to fall down the distribution (a negative). Thus, in part, how higher variance should be viewed depends on value judgements regarding whether such positives outweigh such negatives. These are not simple questions to answer. Also relevant is the underlying explanation for the difference in variance. For example, if the higher variance seen in the first school is a result of its school policies and practices having greater differential effects on different student groups versus the second school, then higher variance might be viewed as a negative as the explanation implies that the school might not be in sufficient control in the implementation of its policies and practices and is exacerbating inequities in student learning versus the first school (Nuttal et al., 1989; Strand, 2010; Scherer & Nilsen, 2019). Though, here too, a tension lies around what is the optimal level of control. Again, these are not simple questions to answer. More generally, school differences in the adjusted variances, just like school differences in the adjusted means, may also reflect unmodelled school differences in student intake and so it is important to attempt to adjust fully for such differences.

A necessary first step to addressing these bigger questions and debates is to first measure school differences in the variance in student adjusted achievement. Only then can school effectiveness and other researchers follow up individual schools which show unusually high or low variance to try to identify the specific school policies and practices which are associated with this. Similarly, only then, can school accountability systems, via school inspections, ask schools to reflect on any unusual school variance scores and discuss these within the broader context of what is happening in these schools and other schools facing similar challenges. All these discussions should be alert to the descriptive rather than causal nature of the statistics and to the limitations of the data more generally, and these statistics should not be used to make automatic high-stakes judgements on schools.

The aim of this article is to therefore broaden the traditional school value-added model to study the effects of schools on not just mean student current achievement, but the variance in student current achievement. We do this by applying mixed-effect location scale (MELS) models to student current achievement. MELS models are an extension to conventional mixed-effects linear regression models that model the residual variance not as a constant, but as a function of the covariates and a new random effect. Thus, the residual variance is now allowed to vary across the schools. Hedeker et al. (2008) illustrated the MELS model in the context of studying intensive longitudinal data on mood. Subsequently, Hedeker and others further developed this class of models and applied it to a range of other longitudinal psychological and health data (e.g., Goldstein et al., 2018; Hedeker et al., 2012; Nordgren et al., 2019; Parker et al., 2021; Rast et al., 2012). Just as mixed-effects models more generally are routinely also applied to clustered cross-sectional data, so can MELS models. Indeed, several such applications have now been published, including in social science research, (Brunton-Smith et al., 2017, 2018; Leckie et al., 2014; McNeish, 2020). However, the applicability of MELS models to school value-added studies has not yet been explored. We address this via an application to school value-added models for school accountability in London, England. Specifically, we examine the following research question: *How does the variance in student adjusted achievement vary across schools?*

This article proceeds as follows. In Section 2, we introduce our application. In Section 3, we present the traditional random-intercept and -slope linear regression school value-added models and their extensions to MELS models. In Section 4, we present the results. In Section 5, we provide a general discussion, including implications of our work for research and school accountability.

## 2. Application

### Background

In England, since 2004, the Government has published school value-added scores derived from school value-added models for all secondary schools in the country in annual school performance tables (https://www.gov.uk/school-performance-tables). These scores aim to measure the value that each school adds to student achievement between the end of primary schooling national Key Stage 2 (KS2) tests (age 11, academic year 6) and the end of compulsory secondary schooling General Certificate of Secondary Education (GCSE) examinations (age 16, academic year 11). The scores play a pivotal role in the national school accountability system, informing school inspections and judgements on schools. They are also promoted to parents as a source of information when choosing schools for their children. Their high stakes use and public presentation have drawn sustained criticism from the academic literature (Goldstein & Speigelhalter, 1996; Leckie & Goldstein, 2009, 2017, 2019; Prior, Jerrim, et al., 2021). Nevertheless, these authors also argue that when used carefully and collaboratively with schools in a sensitive and less public manner there is still an important role for these scores to help identify and understand differences in student outcomes across schools and it is in this spirit that we have carried out the current research (Goldstein, 2020).

## Data, Sample, Variables

We focus on schools in London and on those students who took their GCSE examinations in 2018 and therefore KS2 tests in 2013. The sample is drawn from the National Pupil Database (DfE, 2023) a census of all students in state education and consists of 71,321 students in 465 schools (mean = 153 students per school, range = 14 to 330).

Student current and prior achievement are measured by students' GCSE examination and KS2 test scores (DfE, 2020). We standardize these scores to have means of 0 and SDs of 1 so that the measures can be interpreted in SD units. Henceforth, we refer to these standardized scores simply as the student age 16 and 11 scores. Figure 1 shows both scores are approximately normally distributed and linearly related with a strong Pearson correlation of 0.72. There are very slight floor and ceiling effects in age 16 scores.

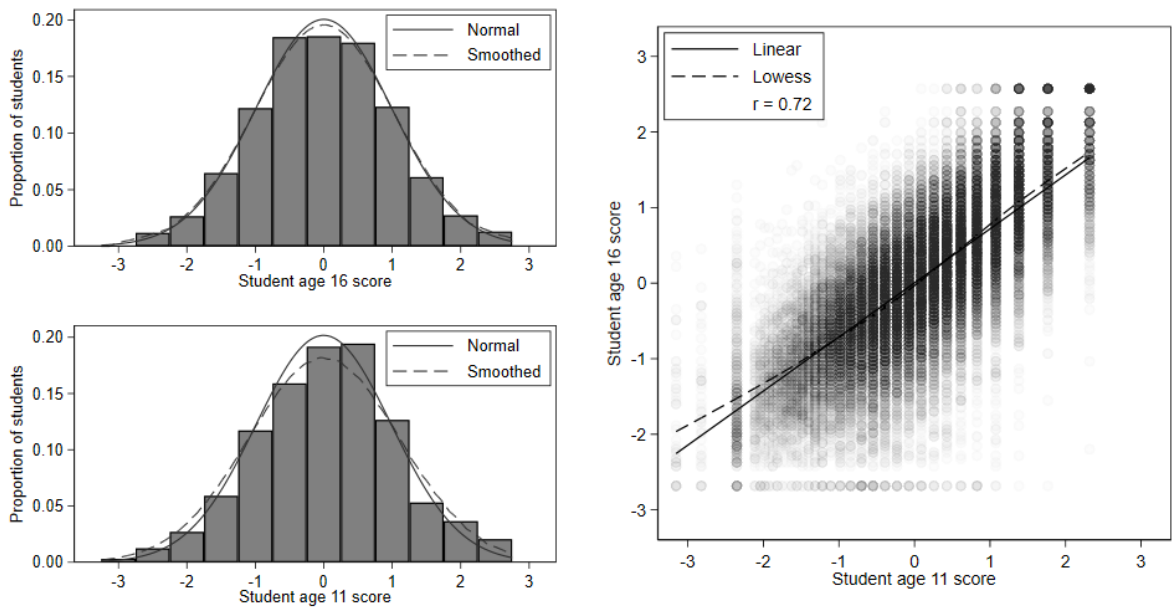

FIGURE 1. *Histograms and scatterplot of student age 16 and age 11 scores.*

Table 1 presents summary statistics for the student characteristics. Of note, 61% of students are non-white and 35% poor (as measured by receipt of free school meals). The London sample is therefore more ethnically diverse and poorer than the full English sample where only around 25% of students are non-white and 25% poor (Leckie & Goldstein, 2019).

TABLE 1

*Summary statistics for the student characteristics (n = 71,321)*

|  | N | % |
| --- | --- | --- |
| Age |  |  |
|   Not summer born | 52,957 | 74.3 |
|   Summer born | 18,364 | 25.8 |
| Gender |  |  |
|   Boy | 35,338 | 49.6 |
|   Girl | 35,983 | 50.5 |
| Ethnicity |  |  |
|   White | 28,070 | 39.4 |
|   Black | 15,633 | 21.9 |
|   Asian | 14,987 | 21.0 |
|   Chinese | 447 | 0.6 |
|   Mixed | 5,795 | 8.1 |
|   Other | 6,389 | 9.0 |
| Language |  |  |
|   English | 42,789 | 60.0 |
|   Not English | 28,532 | 40.0 |
| Special educational needs (SEN) |  |  |
|   Not SEN | 61,189 | 85.8 |
|   SEN | 10,132 | 14.2 |
| Free school meal (FSM) |  |  |
|   Not FSM | 46,500 | 65.2 |
|   FSM | 24,821 | 34.8 |

Table 2 presents summary statistics for the school characteristics. A range of school types operate in London (Leckie & Goldstein, 2019) and we have categorized these into four groups: standard, sponsored academy, converter academy, other. Standard school type encompasses community, foundation, voluntary aided, voluntary controlled, and city technology colleges. In contrast to standard and other schools, academies receive their funding directly from the government rather than through local authorities (school districts). Sponsored academies are mostly underperforming schools which have been required to change to academy status and are run by sponsors. Converter academies are successfully performing schools that have opted to convert to academy status. Other school type encompasses free, studio, university technology colleges (UTCS), and further education colleges. These are more technically or vocationally oriented schools.

A minority of local authorities operate selective rather than comprehensive admissions. In these areas, grammar schools select students based on high performance in entrance examinations and so by definition have high mean age 11 scores and tend also to be educationally advantaged and homogenous in terms of student sociodemographic characteristics. Secondary modern schools take those students not admitted to grammar schools.

TABLE 2

*Summary statistics for the school characteristics ($n = 465$)*

|  | n | % |
|---|---|---|
| Type |  |  |
|   Standard | 151 | 32.5 |
|   Sponsored academy | 93 | 20.0 |
|   Converter academy | 184 | 39.6 |
|   Other | 37 | 8.0 |
| Admissions |  |  |
|   Comprehensive | 425 | 91.4 |
|   Grammar | 19 | 4.1 |
|   Secondary modern | 21 | 4.5 |
| School gender |  |  |
|   Mixed | 340 | 73.1 |
|   Boys | 50 | 10.8 |
|   Girls | 75 | 16.1 |
| Religious |  |  |
|   No | 349 | 75.1 |
|   Yes | 116 | 25.0 |

**3. Models**

**Model 1: Random-intercept Model**

The traditional school value-added model (Aitkin & Longford, 1986; Goldstein et al., 1993; Raudenbush & Bryk, 1986) can be written as the following random-intercept linear regression

$$y_{ij} = \beta_0 + \beta_1 x_{1ij} + u_j + e_{ij} \tag{1}$$

$$u_j \sim N(0, \sigma_u^2)$$

$$e_{ij} \sim N(0, \sigma_e^2)$$

where $y_{ij}$ and $x_{1ij}$ denote current and prior achievement for student $i$ ($i = 1, \ldots, n_j$) in school $j$ ($j = 1, \ldots, J$), $\beta_0$ and $\beta_1$ denote the regression coefficients, $u_j$ the school random intercept effect, $e_{ij}$ the student residual, and where $u_j$ and $e_{ij}$ are assumed independent of one another, independent of $x_{1ij}$, and normally distributed with zero means and constant variances $\sigma_u^2$ and $\sigma_e^2$. As discussed in the Introduction, the independence assumptions are unlikely to hold and so in this article we interpret the school value-added model and the predicted school effects as descriptive rather than causal. Further student and school covariates may be added to this model and we will explore this in the Results section.

The total residual $u_j + e_{ij}$ measures covariate-adjusted (residualized) student current achievement. That is student current achievement having adjusted for prior achievement. The overall average adjusted achievement is 0. The random effect $u_j$ therefore measures the mean student adjusted achievement in each school (the traditional school value-added score) while the residual $e_{ij}$ measures the adjusted achievement of each student relative to their school mean. The random effect variance $\sigma_u^2$ measures the variation in school mean adjusted achievement across schools. The residual variance $\sigma_e^2$ measures the average variance in student adjusted achievement within schools. Crucially, this parameter is averaged across all schools. Thus, while the model allows mean student adjusted achievement to vary from school to school $u_j$, it assumes the variance in student adjusted achievement is the same in every school $\sigma_e^2$ (homoskedasticity).

Figure 2 illustrates the main details of this and subsequent models using hypothetical data on two schools. In each case, $y_{ij}$ is plotted against $x_{1ij}$. In Model 1 (Figure 2A), the two solid lines represent the school-specific relationships $\beta_0 + \beta_1 x_{1ij} + u_j$. The dotted line depicts the average relationship between the two variables $\beta_0 + \beta_1 x_{1ij}$. The school lines are parallel to the average line, because in this model only the intercept $u_j$ differs between schools. The line for school 1 lies above the average line, while the line for school 2 lies below it. The vertical deviations of the school lines from the average line correspond to the school-specific $u_j$. In the current example we have $u_1 > 0 > u_2$. Thus, on average, students in school 1 are predicted to score higher compared to students with the same prior achievement in the average school, while students in school 2 are predicted to score lower. The variability in these mean deviations across all schools corresponds to $\sigma_u^2$. The vertical deviation of the student current achievement scores from their relevant school line corresponds to the $e_{ij}$. The variability in these deviations corresponds to $\sigma_e^2$. This is constant across $x_{1ij}$ and constant across schools.

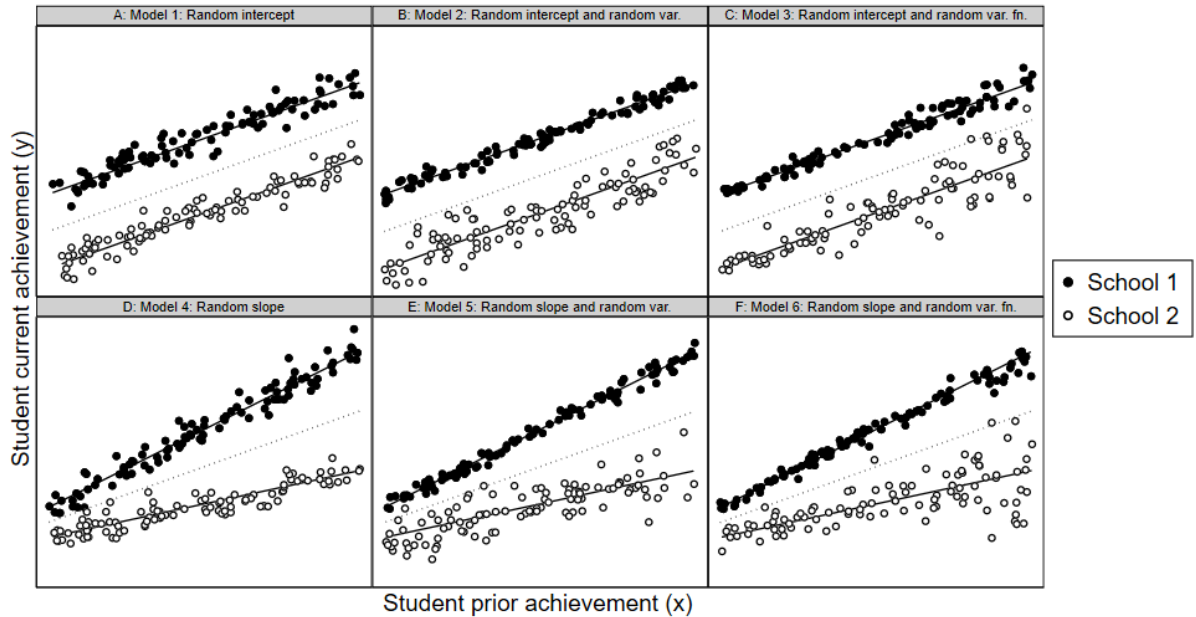

FIGURE 2. *Illustration of different models using hypothetical student current and prior achievement scores data for two schools, school 1 (solid markers) and school 2 (hollow markers). Panel A: Random-intercept model. Panel B: Random-intercept model with random residual variance. Panel C: Random-intercept model with random residual variance function. Panel D: Random-slope model. Panel E: Random-slope model with random residual variance. Panel F: Random-slope model with random residual variance function.*

## Model 2: Random-intercept Model with Random Residual Variance

Model 2 extends Model 1 by allowing the variance in student adjusted achievement $\sigma_e^2$ to vary across schools. We do this by specifying a MELS version of the previous model (Hedeker et al., 2008). The model can be written as

$$y_{ij} = \beta_0 + \beta_1 x_{1ij} + u_j + e_{ij} \tag{2}$$

$$\ln\left(\sigma_{e,j}^2\right) = \alpha_0 + v_j$$

$$\begin{pmatrix} u_j \\ v_j \end{pmatrix} \sim N \left\{ \begin{pmatrix} 0 \\ 0 \end{pmatrix}, \begin{pmatrix} \sigma_u^2 & \\ \sigma_{uv} & \sigma_v^2 \end{pmatrix} \right\}$$

$$e_{ij} \sim N\big(0, \sigma_{e,j}^2\big)$$

where the second line of the equation specifies the residual variance $\sigma_{e,j}^2$ as a log-linear function $\ln(\cdot)$ of a new intercept $\alpha_0$ and a new random school effect $v_j$. The $u_j$ and $v_j$ are assumed bivariate normally distributed and independent of the residuals and covariates. The variance function random intercept variance $\sigma_v^2$ measures the variation in the log of the residual variance across schools. The random intercept-slope covariance $\sigma_{uv}$ measures how $u_j$ and $v_j$ covary. All other terms are defined as before. The log-linear link function ensures the resulting school-specific residual variances $\sigma_{e,j}^2$ and therefore school variances of student adjusted achievement are positive (Hedeker, 2008). Figure 2B illustrates Model 2 where $v_2 > v_1$ and so School 2 shows greater variance in their student adjusted achievements than is the case for School 1 $\sigma_{e,2}^2 > \sigma_{e,1}^2$.

**Model 3: Random-intercept Model with Random Residual Variance Function**

Recall the reason for entering student prior achievement (and potentially further student covariates) into the mean function of the model is that schools should not be held accountable for pre-existing differences in student achievement across schools at the start of the value-added period (Ballou et al., 2004; Leckie & Goldstein, 2019; Leckie & Prior, 2022; Levy et al., 2023). A similar argument applies when comparing the variance in student adjusted achievement across schools. For example, suppose the residual variance increases with increasing student prior achievement. This would suggest that schools with higher mean student prior achievement would in general be expected to show more variable student adjusted achievement than schools with lower mean student prior achievement and this is even though we have adjusted for student prior achievement in the mean function. However, following the arguments underpinning the traditional value-added model, this should be viewed as a reflection of their school intake rather than reflecting their school policies and practices. By entering student prior achievement into the model for the variance, we adjust for this overall variance trend. Focus then shifts to how schools deviate from this overall trend.

Model 3 therefore extends Model 2 by adding student prior achievement to the residual variance function. The model is written as

$$y_{ij} = \beta_0 + \beta_1 x_{1ij} + u_j + e_{ij} \tag{3}$$

$$\ln(\sigma_{e,ij}^2) = \alpha_0 + \alpha_1 x_{1ij} + v_j$$

$$\binom{u_j}{v_j} \sim N\left\{\binom{0}{0}, \begin{pmatrix} \sigma_u^2 & \\ \sigma_{uv} & \sigma_v^2 \end{pmatrix}\right\}$$

$$e_{ij} \sim N\left(0, \sigma_{e,ij}^2\right)$$

where $\alpha_1$ is the residual variance function regression coefficient on $x_{1ij}$. All other terms are defined as before. Where further student and school covariates are added to the mean function, all or a subset of these may also be added to the residual variance function. However, in order to compare school intake-adjusted values of the school variance across schools we must now calculate the residual variance in each school at a common value of $x_{1ij}$ such as the mean. For example, $\sigma_{e,j}^2 = \exp(\alpha_0 + \alpha_1 \bar{x}_{1..} + v_j)$ where $\bar{x}_{1..}$ denotes the mean value for $x_{1ij}$ across all students and schools. Figure 2C illustrates Model 3 where $\alpha_1 > 0$ and so the vertical scatter in student current achievement around each school line increases with student prior achievement in both schools and this is in addition to school 2 continuing to have greater within-school variance than school 1 ($v_2 > v_1$).

**Model 4: Random-slope Model**

Model 4 is the differential effects version (Nuttal et al., 1989; Strand, 2010; Scherer & Nilsen, 2019) of the traditional school value-added model (Model 1) and can be written as the following random-slope linear regression

$$y_{ij} = \beta_0 + \beta_1 x_{1ij} + u_{0j} + u_{1j} x_{1ij} + e_{ij} \tag{4}$$

$$\begin{pmatrix} u_{0j} \\ u_{1j} \end{pmatrix} \sim N \left\{ \begin{pmatrix} 0 \\ 0 \end{pmatrix}, \begin{pmatrix} \sigma_{u0}^2 & \\ \sigma_{u01} & \sigma_{u1}^2 \end{pmatrix} \right\}$$

$$e_{ij} \sim N(0, \sigma_e^2)$$

where $u_{0j}$ and $u_{1j}$ denote the random intercept and random slope effects, assumed bivariate normally distributed and independent of the residual and covariates. The random intercept variance $\sigma_{u0}^2$ measures the variation in school mean adjusted achievement across schools when $x_{1ij} = 0$. The random slope variance $\sigma_{u1}^2$ measures the variation in the slope adjustment for prior achievement across schools. The random intercept-slope covariance $\sigma_{u01}$ measures how these two terms covary. All other terms are defined as before. Where the model includes further student covariates, their regression coefficients may also be allowed to vary across schools.

The total residual, now $u_{0j} + u_{1j} x_{1ij} + e_{ij}$, again measures covariate-adjusted student current achievement. However, school mean student adjusted achievement $u_{0j} + u_{1j} x_{1ij}$ now varies not only across schools, but also across students as a function of the covariate with the random slope $x_{1ij}$. Thus, this version of the model allows schools to be potentially more or less effective for students as a function of their prior achievement.

Figure 2D illustrates Model 4 where $u_{1,1} > u_{1,2}$ and so school 1 shows a steeper regression line than the average line, while school 2 shows a shallower line. The school lines are given by $\beta_0 + \beta_1 x_{1ij} + u_{0j} + u_{1j} x_{1ij}$. The vertical deviations of each school line from the average line correspond to $u_{0j} + u_{1j}x_{1ij}$ and so are a linear function of $x_{1ij}$. The figure shows the school value-added score for school 1 is positive in general, but especially positive for students with high $x_{1ij}$. In contrast, the school value-added score for school 2 is negative in general, but especially negative for students with high $x_{1ij}$.

School mean student adjusted achievement, averaging over all students in each school, is given by $u_{0j} + u_{1j}\bar{x}_{1.j}$ where $\bar{x}_{1.j}$ denotes the average of $x_{1ij}$ in school $j$. For the purpose of comparing schools in terms of their means, it is necessary to evaluate this quantity at common values of $\bar{x}_{1.j}$ for all schools. The variance in student adjusted achievement in each school (over all students) is given by $u_{1j}^2 \text{Var}_j(x_{1ij}) + \sigma_e^2$ where $\text{Var}_j(x_{1ij})$ denotes the variance of $x_{1ij}$ in school $j$. The first component of this expression $u_{1j}^2 \text{Var}_j(x_{1ij})$ captures the variance in student adjusted achievement attributable to interactions between the school effects $u_{1j}$ and the student prior achievement $x_{1ij}$. The magnitude of this component varies across schools. For the purpose of comparing schools in terms of their variances, it is necessary to evaluate this component at a common value of $\text{Var}_j(x_{1ij})$ for all schools, for example the average within school variance of $x_{1ij}$. The second component $\sigma_e^2$ is attributable to all other sources of variance in student adjusted achievement. Crucially, this continues to be assumed constant across schools (homoskedasticity). Thus, adding random slopes only partially recognizes that the variance in student adjusted achievement varies across schools.

**Model 5: Random-slope Model with Random Residual Variance**

Model 5 extends Model 4 by allowing the variance in student adjusted achievement to vary across schools. (Equally Model 5 extends Model 2 by adding a random slope in the mean function to student prior achievement.) We do this by specifying a MELS version of the previous model. The model can be written as

$$y_{ij} = \beta_0 + \beta_1 x_{1ij} + u_{0j} + u_{1j} x_{1ij} + e_{ij} \tag{5}$$

$$\ln\left(\sigma_{e,j}^2\right) = \alpha_0 + v_j$$

$$\begin{pmatrix} u_{0j} \\ u_{1j} \\ v_j \end{pmatrix} \sim N \left\{ \begin{pmatrix} 0 \\ 0 \\ 0 \end{pmatrix}, \begin{pmatrix} \sigma_{u0}^2 & & \\ \sigma_{u01} & \sigma_{u1}^2 & \\ \sigma_{u0v} & \sigma_{u1v} & \sigma_v^2 \end{pmatrix} \right\}$$

$$e_{ij} \sim N\left(0, \sigma_{e,j}^2\right)$$

where the second line of the equation specifies the log-linear function for the residual variance (see also Model 2). The three random effects $u_{0j}$, $u_{1j}$, $v_j$ are assumed trivariate normally distributed and independent of the residuals and covariates. Figure 2E illustrates Model 5 where $v_2 > v_1$ and so School 2 shows greater variance in their student adjusted achievements than is the case for School 1 $\sigma_{e,2}^2 > \sigma_{e,1}^2$ as well as a shallower slope (due to $u_{1,1} > u_{1,2}$).

School mean student adjusted achievement (averaging over all students) is then given by $u_{0j} + u_{1j} \bar{x}_{1.j}$ as it was in the constant residual variance case (Model 4) and so we will again need to evaluate this at a common value of $\bar{x}_{1.j}$ for all schools. The variance in student adjusted achievement in each school (over all their students) is now given by $u_{1j}^2 \text{Var}_j\left(x_{1ij}\right) + \sigma_{e,j}^2$ and so differs from the constant residual variance case (Model 4) in that the last term also now varies across schools.

**Model 6: Random-slope Model with Random Residual Variance Function**

Model 6 extends Model 5 by adding student prior achievement to the residual variance function. (Equally Model 6 extends Model 3 by adding a random slope to student prior achievement.) The model is written as

$$y_{ij} = \beta_0 + \beta_1 x_{1ij} + u_{0j} + u_{1j} x_{1ij} + e_{ij} \tag{6}$$

$$\ln\left(\sigma_{e,ij}^2\right) = \alpha_0 + \alpha_1 x_{1ij} + v_j$$

$$\begin{pmatrix} u_{0j} \\ u_{1j} \\ v_j \end{pmatrix} \sim N \left\{ \begin{pmatrix} 0 \\ 0 \\ 0 \end{pmatrix}, \begin{pmatrix} \sigma_{u0}^2 & & \\ \sigma_{u01} & \sigma_{u1}^2 & \\ \sigma_{u0v} & \sigma_{u1v} & \sigma_v^2 \end{pmatrix} \right\}$$

$$e_{ij} \sim N\big(0, \sigma_{e,ij}^2\big)$$

where $\alpha_1$ is the residual variance function regression coefficient on $x_{1ij}$ (see also model 3). All other terms are defined as before. Figure 2F illustrates Model 6 where $\alpha_1 > 0$ and so the vertical scatter in student current achievement around each school line increases with student prior achievement and this is in addition to school 2 continuing to have a shallower slope ($u_{1,1} > u_{1,2}$) and greater within-school variance than school 1 ($v_2 > v_1$).

As in Model 5 (and Model 4) school mean student adjusted achievement (averaging over all students) is once again given by $u_{0j} + u_{1j}\bar{x}_{1.j}$, while the variance in student adjusted achievement in each school (over all students) is now given by $u_{1j}^2 \text{Var}_j(x_{1ij}) + \text{E}_j\left(\sigma_{e_{ij}}^2\right)$, where $\text{E}_j\left(\sigma_{e_{ij}}^2\right)$ is the mean of the student specific residual variances in school $j$. Crucially, this mean is free to vary across schools.

**Software**

The traditional school value-added models (Model 1 and 4) are typically fitted via maximum likelihood estimation using conventional mixed-effects linear regression routines in standard software (R, SAS, SPSS, Stata). However, the MELS versions of these models (Models 2, 3, 5, 6) cannot be fitted using these routines, nor can they be fitted in specialized mixed-effects modeling packages (HLM, MLwiN). Hedeker and colleagues have developed the MixWILD software to fit MELS models by maximum likelihood estimation (Dzubur et al., 2020). These models can also be fitted via Markov Chain Monte Carlo (MCMC) methods in Stata, and Mplus (McNeish, 2020), as well as dedicated Bayesian software such as Stan (including via the brms package in R; e.g., Parker et al., 2021), WinBUGS, OpenBUGS, and JAGS (including via the R2jags package R: e.g., Barrett et al., 2019). To support readers wishing to implement these models, we present annotated MixWILD, R, and Stata instructions and syntax and simulated data (Section S4 of the Supplemental information).

We fit all models using Stata (StataCorp, 2021). Specifically, we use the bayesmh command which implements an adaptive Metropolis-Hastings MCMC algorithm. We use hierarchical centering reparameterizations to improve mixing. We specify vague (diffuse) normal priors for all regression coefficients and minimally informative inverse Wishart priors for the random effects variance-covariance matrices. We specify overdispersed initial values for all parameters. We fit all models with four chains, each with 5,000 burnin iterations and 10,000 monitoring iterations. We judge convergence using Gelman-Rubin convergence diagnostics (Gelman & Rubin, 1992) and trace, autocorrelation, and scatter plots. All models converged and all parameters had effective sample sizes > 400. We compare model fit using the deviance information criterion (DIC) (Spiegelhalter et al., 2002). Smaller values are preferred.

### 4. Results

**Model 1: Random-intercept Model**

Model 1 (Equation 1) is the traditional school value-added model. In other words, the random-intercept model. For simplicity and because not all researchers wish to additionally include student sociodemographics (Leckie & Prior, 2022; Levy et al., 2023) we only adjust for student prior achievement in this and subsequent models 1-6, but we do explore the role of further covariates in models 7 and 8. For the purpose of comparing to subsequent models, we parameterize $\sigma_e^2$ as $\exp(\alpha_0)$.

Table 3 presents the results. The estimated slope coefficient on student age 11 score is $\hat{\beta}_1 = 0.678$, and so a 1 SD difference in age 11 score is associated with a 0.678 SD difference in age 16 score. The estimated residual variance is $\hat{\sigma}_e^2 = \exp(-0.870) = 0.487$. The estimated total variance in student adjusted achievement is $\hat{\sigma}_u^2 + \hat{\sigma}_e^2 = 0.487$ (and so student age 11 scores accounts for 51% of the variation in student age 16 scores $(= 100\{1 - (\hat{\sigma}_u^2 + \hat{\sigma}_e^2)\}$; Snijders & Bosker, 2012). The estimated between-school variance in school mean adjusted achievement is $\hat{\sigma}_u^2 = 0.067$ and so 14% of the total variation in student adjusted achievement $(= 100\,\hat{\sigma}_u^2/(\hat{\sigma}_u^2 + \hat{\sigma}_e^2)$; Snijders & Bosker, 2012) is variation in the schools means. The between-school variance implies a 95% plausible values range (PVR) for the school means of $(-0.52, 0.50) = \hat{\beta}_0 \pm \Phi^{-1}(0.975)\sqrt{\hat{\sigma}_u^2}$ (where $\Phi^{-1}(\cdot)$ denotes the inverse cumulative standard normal distribution; Raudenbush & Bryk, 2002). Thus, students in what would be deemed the most effective schools (operating at the 97.5th percentile of the distribution of all schools) are predicted to score 1.02 SD higher at age 16 than equivalent students in the least effective schools (operating at the 2.5th percentile). In contrast, the estimated student residual variance $\hat{\sigma}_e^2 = \exp(\hat{\alpha}_0) = 0.419$, is assumed constant, naively implying the variance in student adjusted achievement is the same in every school. Plots confirm that the random effect and residual normality assumptions for this and subsequent models are reasonable (Supplemental information).

TABLE 3

*Results for the Random-intercept Models Adjusting Only for Student Prior Achievement*

|  |  | Model 1 | | Model 2 | | Model 3 | |
|---|---|---|---|---|---|---|---|
|  |  | Est. | SE | Est. | SE | Est. | SE |
|  |  | Mean function | | | | | |
| $\beta_0$ | Intercept | -0.011 | 0.012 | -0.011 | 0.013 | -0.011 | 0.012 |
| $\beta_1$ | Age 11 score | 0.678 | 0.003 | 0.679 | 0.003 | 0.679 | 0.003 |
| $\sigma_u^2$ | School intercept effect variance | 0.067 | 0.005 | 0.067 | 0.005 | 0.067 | 0.005 |
|  |  | Residual variance function | | | | | |
| $\alpha_0$ | Intercept | -0.870 | 0.005 | -0.881 | 0.010 | -0.881 | 0.011 |
| $\alpha_1$ | Age 11 score | | | | | 0.029 | 0.006 |
| $\sigma_v^2$ | School intercept effect variance | | | 0.037 | 0.003 | 0.040 | 0.004 |
|  |  | Assoc. between mean and var. fn. random effects | | | | | |
| $\rho_{uv}$ | Intercept effects correlation | | | -0.472 | 0.048 | -0.484 | 0.047 |
|  |  | Fit statistics | | | | | |
|  | DIC | 140803 | | 139831 | | 139796 | |

*Note.*
Est. and SE denote the posterior means and SDs of the parameter chains. DIC denotes the deviance information criterion.

## Model 2: Random-intercept Model with Random Residual Variance

Model 2 (Equation 2) extends the random-intercept model (Model 1, Equation 1) to allow the residual variance and therefore variance in student adjusted achievement to vary across schools. Model 2 shows a reduction in the DIC of 972 points confirming that this variation in variances is statistically significant. The mean function parameter estimates are largely unchanged. The estimated residual variance function intercept and estimated variance of the new school random effect are $\hat{\alpha}_0 = -0.881$ and $\hat{\sigma}_v^2 = 0.037$. The model-implied population-averaged school variance in student adjusted achievement is estimated as $0.422 = \exp\left(\hat{\alpha}_0 + \frac{\hat{\sigma}_v^2}{2}\right)$ (Hedeker et al., 2008), which, as expected, is close to the Model 1 estimate of 0.419. The estimated population 95% PVR of school variances of student adjusted achievement is $(0.28, 0.61) = \exp\left\{\hat{\alpha}_0 \pm \Phi^{-1}(0.975)\sqrt{\hat{\sigma}_v^2}\right\}$. This range is substantial. For example, the estimated difference in student adjusted achievement between students performing at the 97.5[th] and 2.5[th] percentile within the most variable schools $\hat{\sigma}_{e,j}^2 = 0.61$ is 3.05 SD while in the least variable schools $\hat{\sigma}_{e,j}^2 = 0.28$ it is 2.09 SD (Raudenbush & Bryk, 2002).

Figure 3 plots the predicted school means of student adjusted achievement $u_j$ (y-axis) against the predicted school variances $\hat{\sigma}_{e,j}^2 = \exp(\hat{\alpha}_0 + \hat{v}_j)$ (x-axis). The means and variances are posterior mean predictions and so have been shrunk towards their population average values as a function of their sample size (Snijders & Bosker, 2012). The London average values are illustrated by the horizontal and vertical reference lines. The figure visualizes the substantial variation in both school means and variances of student adjusted achievement described above. While the negative correlation is moderate to large $r = -0.54$, having a high school mean by no means guarantees having a low variance. Equally, there are many instances where schools show similar means but noticeably different variances.

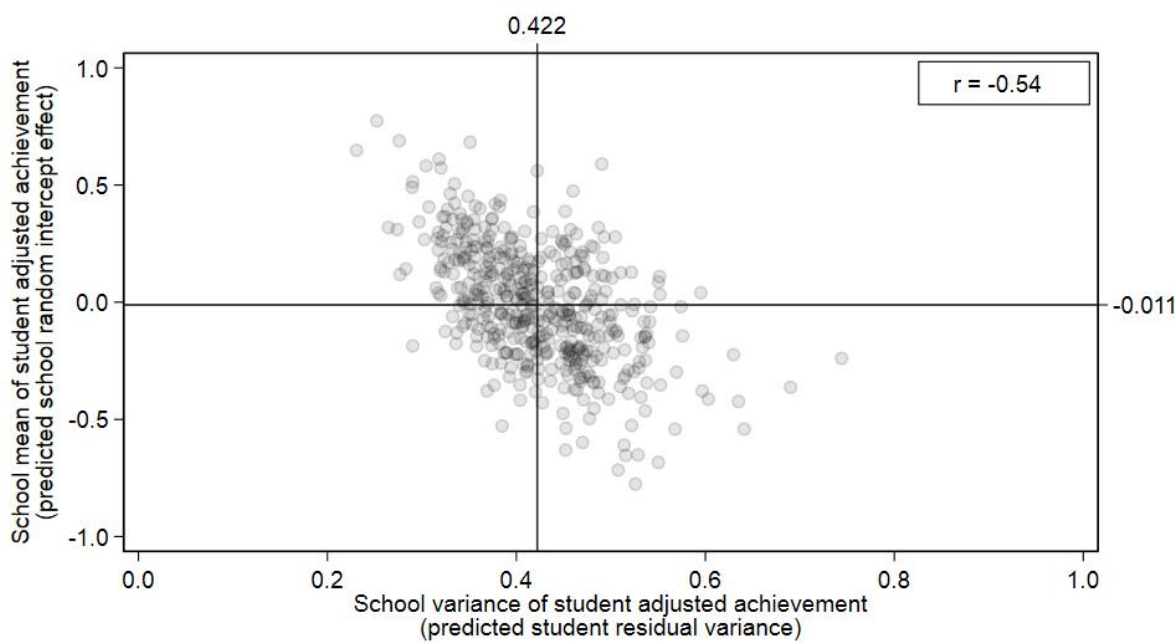

FIGURE 3. *Model 2 scatterplot of school means against school variances of student adjusted achievement. London average values are shown by horizontal and vertical reference lines.*

Figure 4 presents 'caterpillar plots' of the 465 predicted school means (left panel) and school variances (right panel) (Goldstein, 2011). Such plots are routinely used by researchers and accountability systems to identify schools that are significantly different from average (e.g., Prior et al., 2021). The distribution of the school variances is positively skewed, consistent with being modelled as log-normally distributed. Schools with fewer students have wider 95% credible intervals than schools with more students. Only 117 out of the 465 schools (25%) can be statistically separated from the overall average in terms of their school variances compared to 320 schools (69%) when we consider the school means.

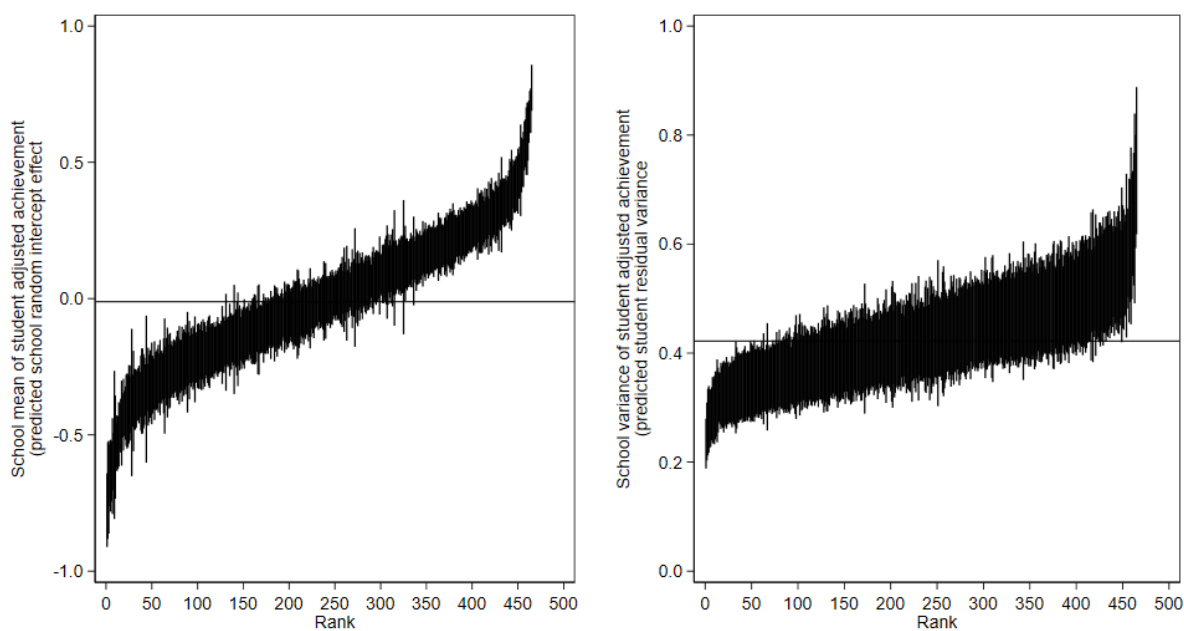

FIGURE 4. *Model 2 caterpillar plots for school means (left) and school variances (right) of student adjusted achievement presented in rank order. Posterior means with 95% credible intervals.*

**Model 3: Random intercept Model with Random Residual Variance Function**

Model 3 (Equation 3) further extends the random-intercept model to allow the residual variance to vary not just across schools (Model 2, Equation 2), but additionally as a function of student prior achievement. Model 3 is preferred to Model 2 ($\Delta\text{DIC} = 34$) showing the residual variance significantly increases with student age 11 scores $\hat{\alpha}_1 = 0.029$. Thus, schools with in general higher age 11 scores are predicted to show higher variance in student adjusted achievement. However, this relationship is very weak. The estimated population 95% PVR of school intake adjusted variances of student adjusted achievement is effectively the same as in the previous model where we did not adjust for school intake, $(0.28, 0.61) = \exp\left\{\hat{\alpha}_0 + \hat{\alpha}_1\bar{\bar{x}}_{1..} \pm \Phi^{-1}(0.975)\sqrt{\hat{\sigma}_v^2}\right\}$ where $\bar{\bar{x}}_{1..} = 0$ denotes the London-wide average value for $x_{1ij}$. That is, the variation in the variance in student adjusted achievement across schools is not simply explained by some schools showing in general higher age 11 scores and therefore higher variances than others.

**Model 4: Random-slope Model**

Model 4 (Equation 4) is the differential effectiveness version of the traditional school-value-added model. In other words, the random-slopes model. Recall that this model, like the traditional random-intercepts model (Model 1, Equation 1), assumes the residual variance is once again constant across all students and schools $\sigma_e^2$. As in Model 1, we parameterize $\sigma_e^2$ as $\exp(\alpha_0)$.

Table 4 presents the results. Model 4 is preferred to Model 1 ($\Delta\text{DIC} = 281$) confirming the age 11 slope varies significantly across schools. The estimated mean and variance of the age 11 slope across schools are $\hat{\beta}_1 = 0.675$ and $\hat{\sigma}_{u1}^2 = 0.004$. The latter implies an estimated 95% PVR of school slopes of $(0.55, 0.80) = \hat{\beta}_1 \pm \Phi^{-1}(0.975)\sqrt{\hat{\sigma}_{u1}^2}$. Figure 5 visualizes this variation for the sample schools by plotting the predicted school lines based on Model 1 (left panel) and Model 4 (right panel). The plots appear very similar suggesting that while the random slopes are statistically significant they are not practically significant. Indeed, moving from Model 1 to Model 4 the residual variance reduces by just 0.70%. Thus, in contrast to the literature which tends to show larger variation in school effects among low prior achievers versus high prior achievers, we find no such pattern. (Nuttal et al., 1989; Strand, 2010; Scherer & Nilsen, 2019).

TABLE 4

*Results for the Random-slope Models Adjusting Only for Student Prior Achievement*

| | | Model 4 | | Model 5 | | Model 6 | |
|---|---|---|---|---|---|---|---|
| | | Est. | SE | Est. | SE | Est. | SE |
| | | Mean function | | | | | |
| $\beta_0$ | Intercept | -0.017 | 0.013 | -0.015 | 0.013 | -0.015 | 0.013 |
| $\beta_1$ | Age 11 score | 0.675 | 0.004 | 0.673 | 0.004 | 0.672 | 0.004 |
| $\sigma_{u0}^2$ | School intercept effect variance | 0.068 | 0.005 | 0.069 | 0.005 | 0.069 | 0.005 |
| $\sigma_{u1}^2$ | School slope effect variance | 0.004 | 0.000 | 0.004 | 0.000 | 0.004 | 0.000 |
| $\rho_{u0u1}$ | Intercept slope effects correlation | 0.278 | 0.064 | 0.231 | 0.066 | 0.229 | 0.067 |
| | | Residual variance function | | | | | |
| $\alpha_0$ | Intercept | -0.877 | 0.005 | -0.889 | 0.010 | -0.889 | 0.011 |
| $\alpha_1$ | Age 11 score | | | | | 0.036 | 0.006 |
| $\sigma_v^2$ | School intercept effect variance | | | 0.037 | 0.003 | 0.040 | 0.004 |
| | | Assoc. between mean and var. fn. random effects | | | | | |
| $\rho_{u0v}$ | Intercept effects correlation | | | -0.476 | 0.048 | -0.494 | 0.047 |
| $\rho_{u1v}$ | Slope intercept effect correlation | | | -0.089 | 0.075 | -0.111 | 0.076 |
| | | Fit statistics | | | | | |
| | DIC | 140522 | | 139546 | | 139495 | |

*Note.*
Est. and SE denote the posterior means and SDs of the parameter chains. DIC denotes the deviance information criterion.

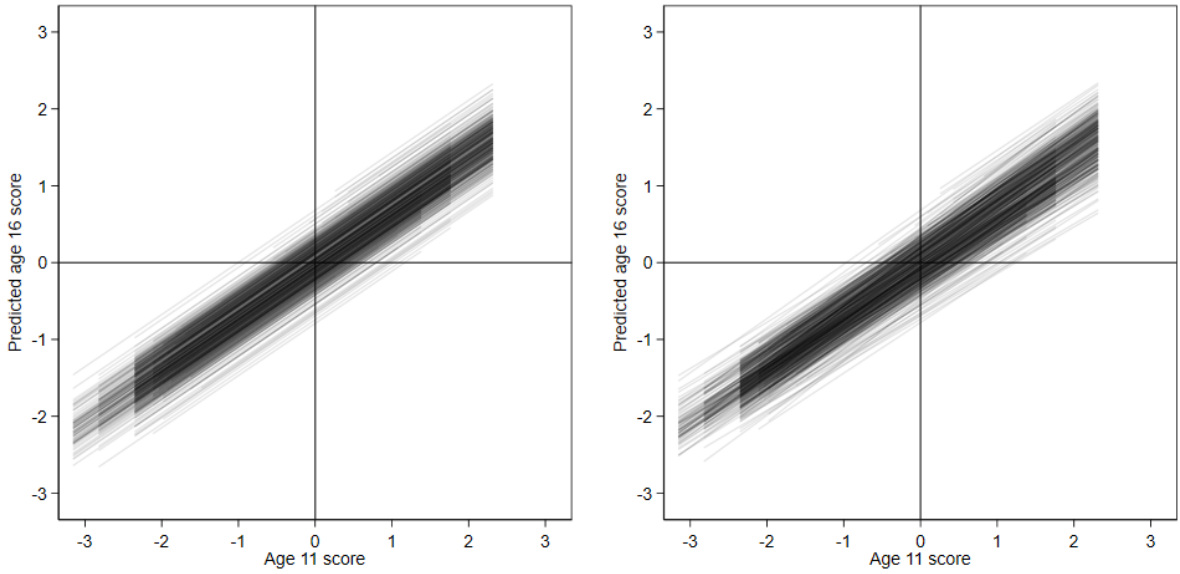

FIGURE 5. *Model 1 and Model 4 school regression lines of predicted age 16 scores against age 11 scores for random-intercept model (left) and random-slope model (right).*

**Model 5: Random-slope Model with Random Residual Variance**

Model 5 (Equation 5) extends the random-slope model (Model 4, Equation 4) to allow the residual variance to vary across schools. Thus, the move from Model 4 to 5 for the current random-slope model mirrors the move we explored from Model 1 to 2 for the earlier random-intercept versions of these models.

Model 5 allows us to quantify the relative importance of the differential school effects with respect to prior achievement as a component of the overall variance in student adjusted achievement in each school. We calculate the estimated variance for each school in our sample for a common reference distribution of students with student age 11 score variance $\bar{s}^2_{x_{1..}} = 0.83$ (the mean of the sample school variances of student prior achievement). The resulting expression is $\hat{u}^2_{1j}\bar{s}^2_{x_{1..}} + \hat{\sigma}^2_{e,j}$ where $\hat{\sigma}^2_{e,j} = \exp(\hat{\alpha}_0 + \hat{v}_j)$ (see Section 3, Model 5). The first component $\hat{u}_{1j}^2 \bar{s}_{x_{1\cdot}}^2$ gives the variance attributable to the random slope interactions $u_{1j}x_{1ij}$. The second component $\hat{\sigma}_{e,j}^2$ captures all remaining variance. The first component is very small accounting for less than 1% of the variance in nearly all schools. In sum, the inclusion of the random slope on prior achievement has done very little to explain the variance in student adjusted achievement in each school.

**Model 6: Random-slope Model with Random Residual Variance Function**

Model 6 (Equation 6) further extends the random-slope model to allow the residual variance to vary not just across schools (Model 5, Equation 5), but additionally as a function of student prior achievement. Thus, the move from Model 5 to 6 for the current random-slope model mirrors the move we explored from Model 2 to 3 for the earlier random-intercept versions of these models. As with the sequence of random intercept models, Model 6 shows the residual variance in the random-slope model significantly increases with student age 11 scores $\hat{\alpha}_1 = 0.036$. However, as with the random-intercept models, this effect is slight and does little to explain the variation in school variances across schools. Given adding the random slope has little practical importance and in order to illustrate the subsequence models as simply as possible, we return to the sequence of random-intercept models.

**Model 7: Random-intercept Model with Random Variance Function and Student Characteristics**

Model 7 extends Model 3 by adding student age, gender, first language, SEN status, and FSM status into the mean and residual variance functions (Table 1). Adding these characteristics to the mean function implies students are now compared to other students across London who not only share the same age 11 score, but who also share the same sociodemographic characteristics. The aim is to ensure that schools do not appear more or less effective simply as a result of recruiting more or less educationally advantaged students (Leckie & Goldstein, 2019). The resulting improved accuracy of the predicted age 16 scores will lead the student adjusted achievement scores to in general reduce in absolute magnitude (and reorder) leading the overall variance in student adjusted achievement to decrease. In turn, the school means and variances of student adjusted achievement scores will also change, again in general reducing in magnitude and reordering. We then further adjust the school variances of student adjusted achievement by including the student characteristics in the student residual variance function. This ensures that if there are any London-wide relationships between the variance in student adjusted achievement and particular student characteristics, this again will not benefit or count against schools with disproportionate numbers of these students.

Table 5 presents the results. Model 7 is preferred to Model 3 ($\Delta$DIC = 7247) confirming the statistical importance of the student characteristics. First consider the mean function. The results show that summer born students, girls, all ethnic minority groups except Mixed ethnicity students (relative to White), and students who speak English as a second language, are all predicted to score higher at age 16, than otherwise equivalent students. SEN and FSM students, in contrast, are predicted to score lower than otherwise equivalent students. These results are established and consistent with the literature (Leckie & Goldstein, 2019). What is not known is whether there are also sociodemographic differences in the variance in student adjusted achievement. The results show that, all else equal, the residual variance and therefore variance in student adjusted achievement again increases with age 11 scores but is now also shown to be higher for SEN and FSM students than for otherwise equal students. Thus, it proves harder to predict reliably the age 16 scores of these student groups relative to other student groups. In contrast, summer born students, girls, Black, and Asian students show lower variance in student adjusted achievement and therefore appear to perform in a more consistent fashion than otherwise equal student groups within schools.

TABLE 5

*Results for the Random-intercept Models Adjusting for Student Prior Achievement and Student and School Characteristics*

| | | Model 7 | | Model 8 | |
|---|---|---|---|---|---|
| | | Est. | SE | Est. | SE |
| | | Mean function | | | |
| $\beta_0$ | Intercept | -0.129 | 0.012 | -0.235 | 0.017 |
| $\beta_1$ | Age 11 score | 0.634 | 0.003 | 0.632 | 0.003 |
| $\beta_2$ | Summer born | 0.045 | 0.005 | 0.044 | 0.005 |
| $\beta_3$ | Girl | 0.219 | 0.005 | 0.218 | 0.005 |
| $\beta_4$ | Ethnicity: Black | 0.015 | 0.006 | 0.014 | 0.007 |
| $\beta_5$ | Ethnicity: Asian | 0.152 | 0.008 | 0.150 | 0.008 |
| $\beta_6$ | Ethnicity: Chinese | 0.296 | 0.028 | 0.290 | 0.028 |
| $\beta_7$ | Ethnicity: Mixed | 0.001 | 0.009 | 0.000 | 0.009 |
| $\beta_8$ | Ethnicity: Other | 0.089 | 0.010 | 0.088 | 0.009 |
| $\beta_9$ | First language not English | 0.162 | 0.006 | 0.162 | 0.006 |
| $\beta_{10}$ | Special educational needs (SEN) | -0.276 | 0.008 | -0.276 | 0.008 |
| $\beta_{11}$ | Free school meal (FSM) | -0.193 | 0.005 | -0.192 | 0.005 |
| $\beta_{12}$ | School type: Sponsored academy | | | 0.055 | 0.025 |
| $\beta_{13}$ | School type: Converter academy | | | 0.082 | 0.020 |
| $\beta_{14}$ | School type: Other | | | 0.023 | 0.038 |
| $\beta_{15}$ | School admissions: Grammar | | | 0.396 | 0.049 |
| $\beta_{16}$ | School admissions: Secondary modern | | | -0.118 | 0.045 |
| $\beta_{17}$ | School gender: Boys | | | 0.053 | 0.032 |
| $\beta_{18}$ | School gender: Girls | | | 0.064 | 0.027 |
| $\beta_{19}$ | School religious | | | 0.139 | 0.022 |
| $\sigma_{u0}^2$ | School intercept effect variance | 0.050 | 0.004 | 0.037 | 0.003 |
| | | Residual variance function | | | |
| $\alpha_0$ | Intercept | -0.948 | 0.015 | -0.889 | 0.024 |
| $\alpha_1$ | Age 11 score | 0.077 | 0.006 | 0.081 | 0.006 |
| $\alpha_2$ | Summer born | -0.044 | 0.012 | -0.045 | 0.012 |
| $\alpha_3$ | Girl | -0.059 | 0.012 | -0.061 | 0.012 |
| $\alpha_4$ | Ethnicity: Black | -0.154 | 0.016 | -0.156 | 0.016 |
| $\alpha_5$ | Ethnicity: Asian | -0.105 | 0.018 | -0.106 | 0.018 |
| $\alpha_6$ | Ethnicity: Chinese | -0.088 | 0.072 | -0.080 | 0.069 |
| $\alpha_7$ | Ethnicity: Mixed | -0.028 | 0.022 | -0.035 | 0.021 |
| $\alpha_8$ | Ethnicity: Other | -0.014 | 0.020 | -0.015 | 0.021 |
| $\alpha_9$ | First language not English | -0.002 | 0.013 | -0.005 | 0.013 |
| $\alpha_{10}$ | Special educational needs (SEN) | 0.204 | 0.016 | 0.203 | 0.016 |
| $\alpha_{11}$ | Free school meal (FSM) | 0.103 | 0.012 | 0.099 | 0.012 |
| $\alpha_{12}$ | School type: Sponsored academy | | | 0.011 | 0.028 |
| $\alpha_{13}$ | School type: Converter academy | | | -0.048 | 0.023 |
| $\alpha_{14}$ | School type: Other | | | 0.053 | 0.042 |
| $\alpha_{15}$ | School admissions: Grammar | | | -0.280 | 0.052 |
| $\alpha_{16}$ | School admissions: Secondary modern | | | -0.068 | 0.044 |
| $\alpha_{17}$ | School gender: Boys | | | 0.002 | 0.034 |
| $\alpha_{18}$ | School gender: Girls | | | 0.015 | 0.029 |
| $\alpha_{19}$ | School religious | | | -0.110 | 0.023 |
| $\sigma_v^2$ | School intercept effect variance | 0.032 | 0.003 | 0.026 | 0.003 |
| | | Association between mean and variance function random effects | | | |
| $\rho_{u0v}$ | Intercept effects correlation | -0.409 | 0.050 | -0.282 | 0.057 |
| | | Fit statistics | | | |
| | Deviance information criterion (DIC) | 132549 | | 132539 | |

*Note.*
Est. and SE denote the posterior means and SDs of the parameter chains. Student ethnicity reference group is White. School type reference group is standard. School admissions reference group is comprehensive. School gender reference group is mixed-sex school.

Figure 6 presents scatterplots of the school means and variances of student adjusted achievement based on the current model which adjusts for student background against those based on Model 3 which ignores student background. The purpose of this figure is to explore the sensitivity of the school means and variances to the additional adjustments for student background and to therefore assess the importance of making such adjustments or not (Leckie & Prior, 2022; Levy et al., 2023). We calculate the estimated school variances in each model by plugging in the sample mean values for the covariates (Table 1) in the residual variance function, and so $\hat{\sigma}_{e,j}^2 = \exp(\hat{\alpha}_0 + \hat{\alpha}_1 0.258 + \cdots + \hat{\alpha}_{11} 0.348 + \hat{v}_j)$. The plots show both the school means and the school variances are correlated 0.94 across the two models. Thus, schools which show high mean adjusted achievement when one ignores student background nearly always still show high mean adjusted achievement after adjustment. The same applies for school variances of student adjusted achievement. However, even with such high correlations, the rank ordering of those schools whose social mix differ most markedly from the London-wide average still change considerably as shown by schools located furthest away from the 45-degree line in the bottom plots. Thus, the decision of whether to adjust for student background has a bearing on the manner in which many individual schools are viewed in terms of their school variances as well as their school means.

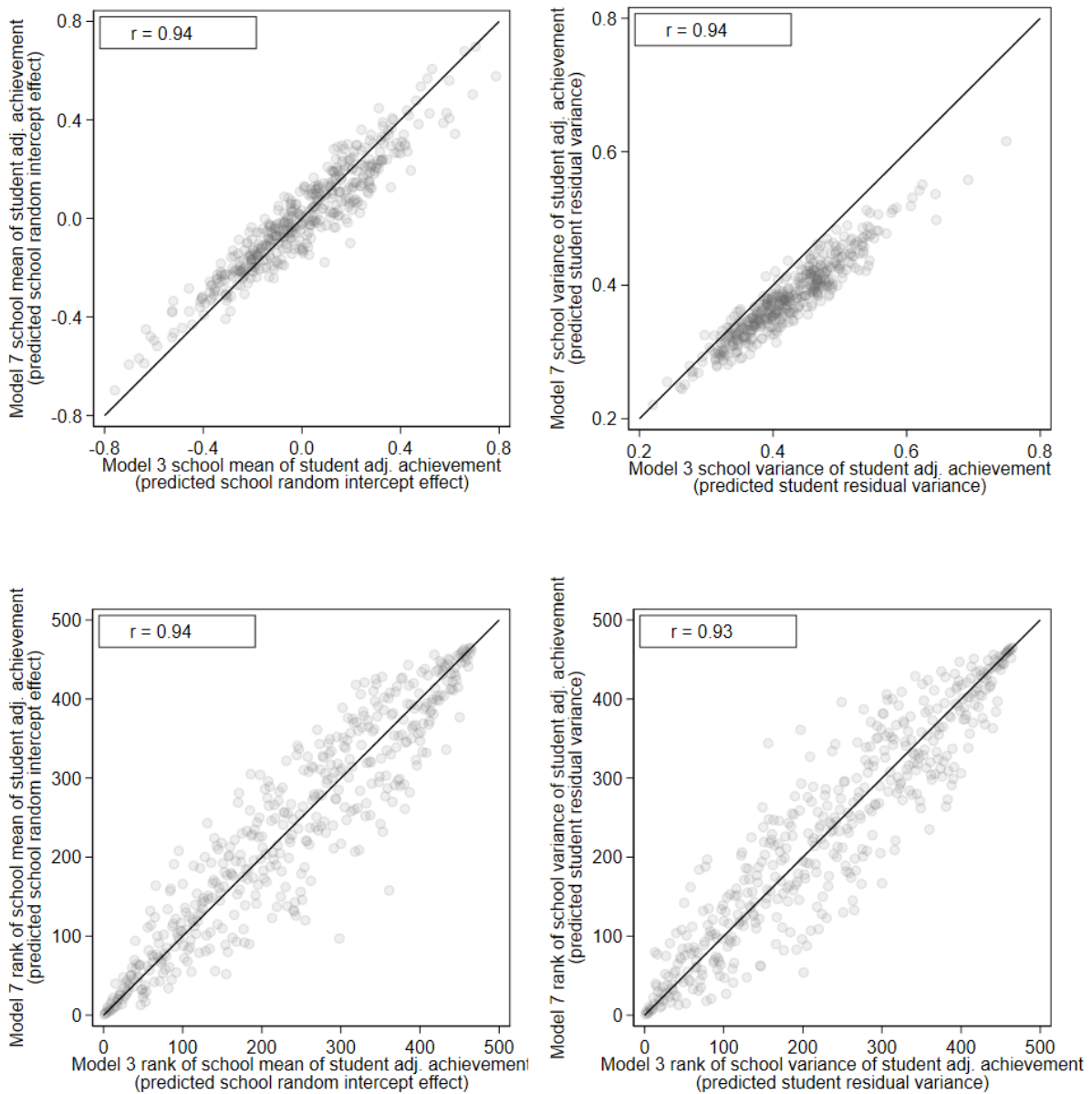

FIGURE 6. *Model 7 against Model 3 scatterplots of school means of student adjusted*

*achievement (top left), school variances of student adjusted achievement (top right), ranks of*

*school means of student adjusted achievement (bottom left), and ranks of school variances of*

*student adjusted achievement (bottom right).*

## Model 8: Random-intercept Model with Random Variance Function and School Characteristics

We now shift from attempting to best define and measure student adjusted achievement, and therefore the school means and variances of student adjusted achievement, to attempting to explain why some schools show higher mean student adjusted achievement and lower variance in student adjusted achievement than others. Unfortunately, we do not observe school policies and practices in our data. However, we do observe some school characteristics (Table 2). Model 8 extends Model 7 by adding school type, school admissions, school gender (mixed, boys, girls), and school religion to the mean and residual variance functions.

The results (Table 5) for the existing mean and residual variance function regression coefficients are very similar to before and so we restrict our interpretation here to the new results. First, consider the mean function. Relative to standard school types, school mean adjusted achievement is somewhat higher in sponsored and converter academies having adjusted for the other covariates. Similarly, school mean adjusted achievement is higher in girls schools and religious schools, all else equal. However, the most sizeable differential is related to school admissions: school mean adjusted achievement is considerably higher in grammar schools and lower in secondary modern schools relative to comprehensive schools. These results agree with the literature (Leckie & Goldstein, 2019). With respect to the residual variance function, we see new findings. School variances in student adjusted achievement tend to be lower in converter academies compared to standard school types, lower in grammar schools versus comprehensive school types, and lower in religious schools versus non-religious schools, and this is after adjusting for London wide relationships between the variance in student adjusted achievement and student characteristics. Thus, students in converter academies, grammar, and religious schools not only tend to show higher student adjusted achievement on average, but also tend to show more consistent student adjusted achievement.

## 5. Discussion

In this article, we have argued that the focus of school value-added models should broaden to measure not just school mean differences in student adjusted achievement (student achievement beyond that predicted by student prior achievement and other student background characteristics), but school variance differences in student adjusted achievement. To study school variance differences, we have proposed extending the traditional school value-added model, a random-intercept mixed-effects linear regression of student current achievement on prior achievement and other student background characteristics, by modeling the residual variance as a log-linear function of the student covariates and a new random school effect. The school random intercept effect and random residual variance in this model measure the school mean and variance in student adjusted achievement. This model can be viewed as an application of the MELS model popular in biostatistics (Hedeker et al., 2008). It is, however, important to reiterate that the school value-added models and their respective predicted school effects should be viewed as descriptive rather than causal since these models do not address the complex selection into schools processes that will be in play in many school systems.

We have illustrated this extended school value-added model with an application to schools in London. In response to our research question: Our results suggest meaningful differences in the variance in student adjusted achievement across schools. We also find a moderate to large negative association between the school mean and variance in student adjusted achievement. Thus, schools which show the highest mean student adjusted achievement also tend to be the schools which show the lowest variance in student adjusted achievement. One process by which school variance differences may arise is if there is a London-wide negative relationship between the variance in student adjusted achievement and student prior achievement. We adjusted for this by entering student prior achievement into the residual variance function. A second process by which school variance differences may arise is via interaction effects between the different school policies and practices envisaged to be represented by the school random intercept effect and observed and unobserved student characteristics. Previous research has studied this via entering a school random slope on student prior achievement and this showed schools to be differentially effective for students with low, middle, and high prior achievement. In our application, however, these school-by-student prior achievement interactions are small and explain little of the variation in school variances between schools. We then turned our attention to entering student characteristics into the model, both in the mean and residual variance functions, to better measure student adjusted achievement. In terms of new results, we find that FSM and SEN students show greater variance in student adjusted achievement and therefore less predictable age 16 scores than otherwise equal students. The resulting predicted school means and variances of student adjusted achievement, however, are similar to those based on the model which only adjusts for student prior achievement. Nevertheless, schools whose sociodemographic student mix differ most from the average school still move up and down the London-wide rankings considerably, demonstrating the importance of adjusting for student background at least for some schools (Leckie and Goldstein, 2019; Leckie and Prior, 2022; Levy et al., 2023). Finally, we shifted our emphasis from measuring school means and variances of student adjusted achievement to seeking to explain them. We find converter academies and grammar schools tend to show lower variances in student adjusted achievement than other school types. Importantly, here too we adjusted for any overall relationship between the variance in student adjusted achievement and student prior achievement and background characteristics and so these differences in school variances lie beyond this simple explanation.

Future studies might seek to identify whether school variance differences can be predicted by specific school policies and practices. It will also be interesting and important to explore the role of school composition covariates such as the school mean and school SD of the student prior achievement (Raudenbush and Bryk, 2002). One issue that such studies should bear in mind is that some student current achievement measures may exhibit floor or ceiling effects. Where these are pronounced, they may bias the model parameters relative to fitting models to measures without such effects. Tobit versions of the models might be considered to address this issue (Lu, 2018). Another issue is sample-size requirements. In general, we found that the residual variance function regression coefficients and predicted school effects were less precisely estimated than their analogous quantities in the mean function. This suggests that larger sample sizes are needed for these models than traditionally used for school value-added studies. Future studies might therefore use power calculations to guide such decisions (Walters et al., 2018).

More generally, however, expanding the focus of school value-added models to consider schools effects on the variance in student achievement raises value judgements and interpretational challenges that future work will need to engage with. Fundamentally, it is not clear how positively or negatively higher or lower variances should be viewed in general. Similarly, where a given school policy or practice is identified as driving school differences in variance via differential effects on students as a function of their observed and unobserved characteristics, it will not typically be clear what the optimal degree of differential impact might be. Even if it is decided that higher variance should be interpreted in a particular way, faced now with two summaries of school effects on student learning (mean and variance effects), researchers and school accountability systems must make further value judgements as to how to best combine them into any overall summary of school effectiveness for the purpose of making overall inferences, judgements and decisions about schools (Prior, Goldstein, et al., 2021). Crucially, it is only by extending the school value-added model to allow for school effects on the variance in student adjusted achievement that such debates are made possible. The extension we have presented paves the way for new substantive research into the reasons behind differences in variability and therefore how best such differences should be interpreted and addressed.

The school value-added model presented here can be further extended in various ways beyond simply adding further covariates and random slopes suggesting avenues for new methodological research. First, in the school effectiveness literature, there is interest in studying the consistency of school effects across academic subjects (Goldstein, 1997; Reynolds et al., 2014; Teddlie & Reynolds, 2000). We can further develop the school value-added model to study this phenomenon with respect to the school variance in student adjusted achievement. Essentially, we would fit a multivariate response version of this model for multiple student achievement scores (Kapur et al., 2015; Leckie, 2018; Pugach et al., 2014). The model would have multiple residual variance functions, one for each academic subject. We can then study the correlations of the school means and variances of student adjusted achievement across subjects. Second, the same multivariate response version of the model can be used to study the stability of school effects over time. Here we would fit a multivariate response model to a single achievement score, but for multiple student cohorts (Leckie & Goldstein, 2009). Third, we could include a random slope in the residual variance function (Goldstein et al., 2018; McNeish, 2021) to study whether schools exacerbate or mitigate any overall relationship between the variance in student adjusted achievement and student prior achievement. Fourth, while we have flexibly modelled the residual variance, we have not modelled the random intercept variance (the random slope model relaxed this, but in a rather specific way). It is also possible to model the random intercept variance as a log-linear function of school covariates (Hedeker et al, 2008). For example, the variability of school mean adjusted achievement scores across schools may appear greater for some school groups than others and this could then be tested by introducing the school group variable as a covariate in this second variance function. Fifth, we can expand the model to three levels to incorporate an additional random effect into the mean and residual variance functions relating to, for example, school district and thereby study school district differences in the mean and variance in student adjusted achievement. This then raises the possibility of entering school district random effects into the school random intercept variance function since school mean adjusted achievement might vary more in some school districts than in others and so with this extension we can potentially study differential school level inequalities in the education system by school district (Leckie and Goldstein, 2015). Alternatively, teacher random effects could be introduced as a new level between the student and school level. Finally, our focus has been on shifting attention from studying school mean of student adjusted achievement to additionally focusing on the variance in student adjusted achievement. In future work it would be interesting to explore further ways the distribution of student adjusted achievement might vary across schools, for example, with respect to skewness.

**Mixed-Effects Location Scale Models for Joint Modeling School Value-Added Effects on**

**the Mean and Variance of Student Achievement**

**Supplementary File**

**S1. Supplementary Figures**

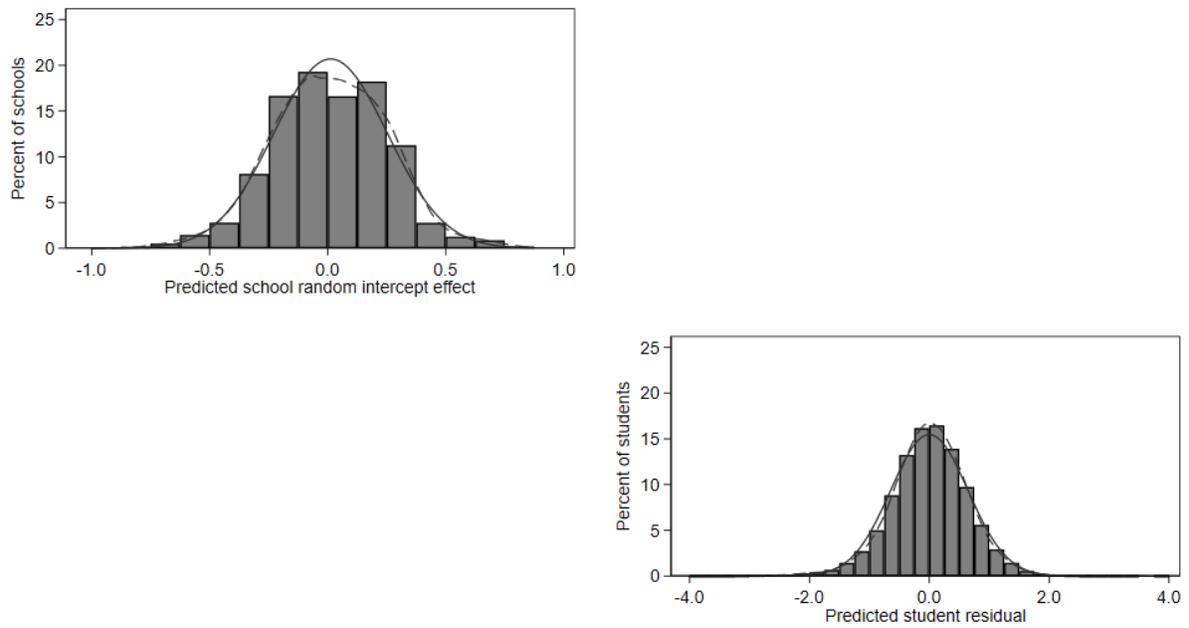

FIGURE S1. *Model 1 histograms of predicted school random intercept effects (top left) and student residuals (bottom right), each with superimposed normal curves (solid) and kernel density curves (dashed).*

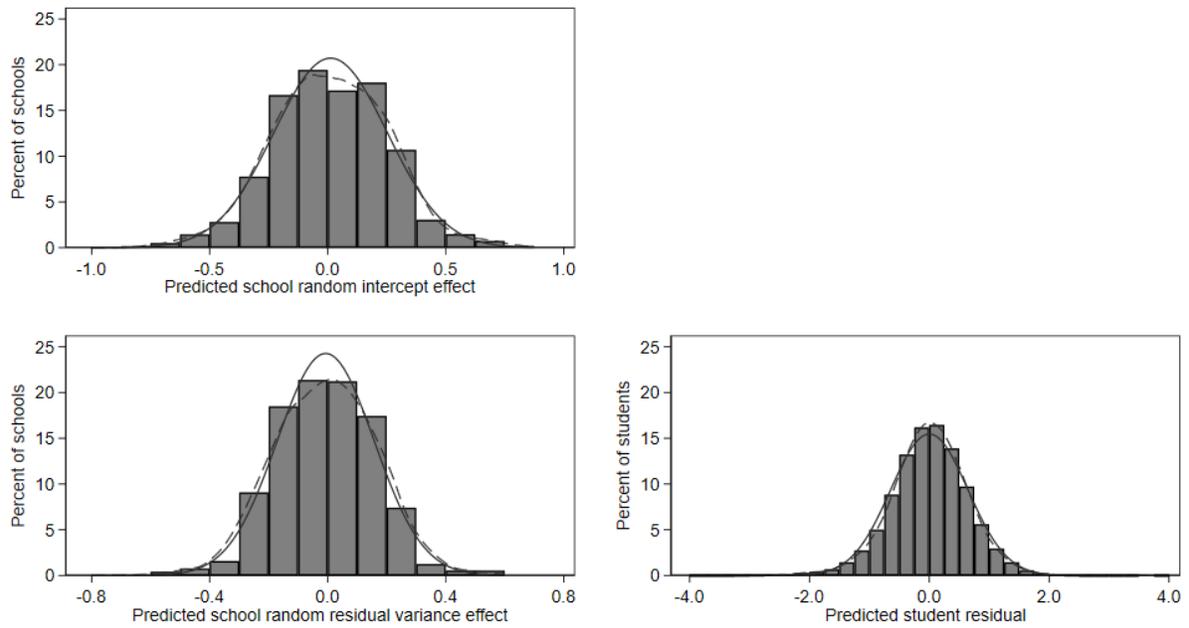

FIGURE S2. *Model 2 histograms of predicted school random intercept effects (top left), school random residual variance effects (bottom left), and student residuals (bottom right), each with superimposed normal curves (solid) and kernel density curves (dashed).*

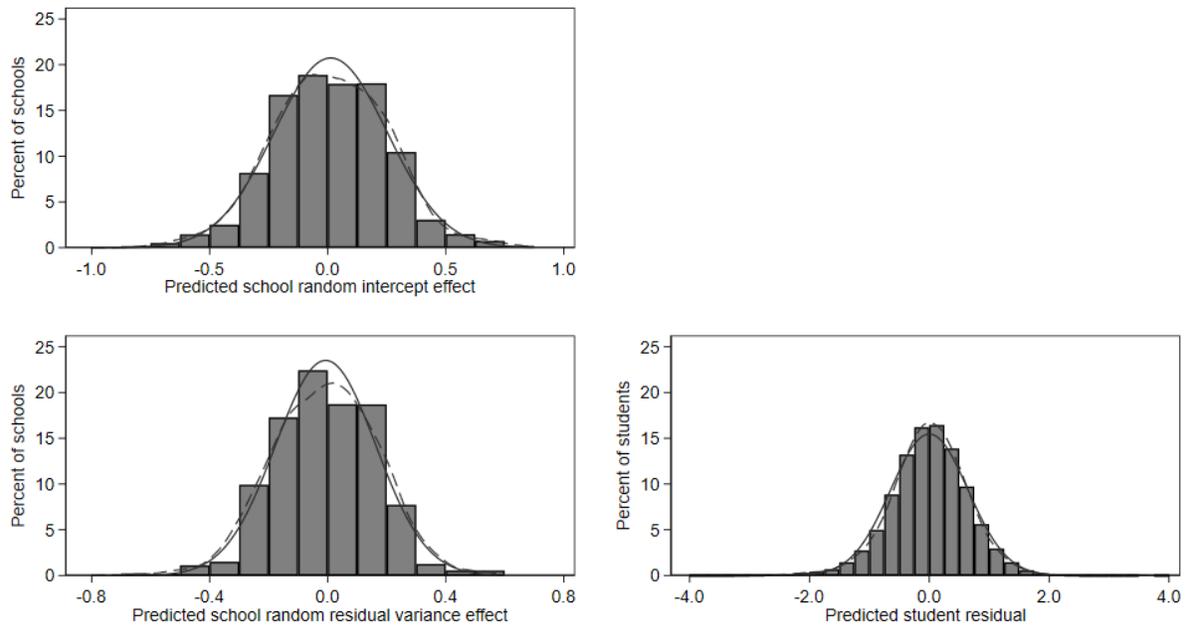

FIGURE S3. *Model 3 histograms of predicted school random intercept effects (top left), school random residual variance effects (bottom left), and student residuals (bottom right), each with superimposed normal curves (solid) and kernel density curves (dashed).*

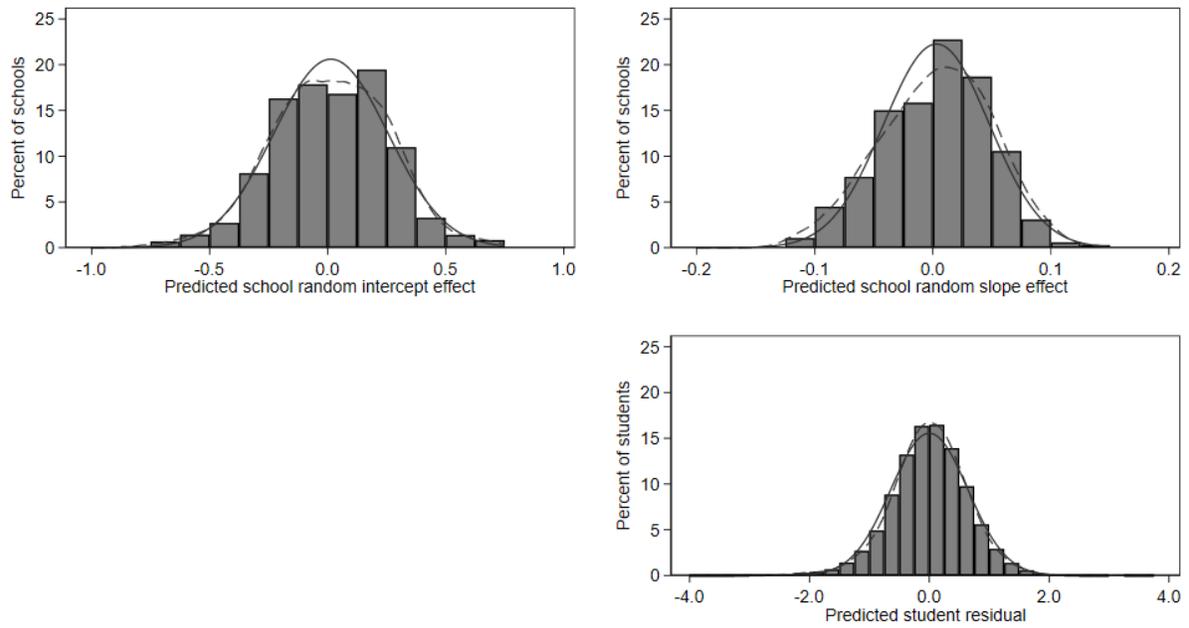

**FIGURE S4.** *Model 4 histograms of predicted school random intercept effects (top left), school random slope effects (top right), and student residuals (bottom right), each with superimposed normal curves (solid) and kernel density curves (dashed).*

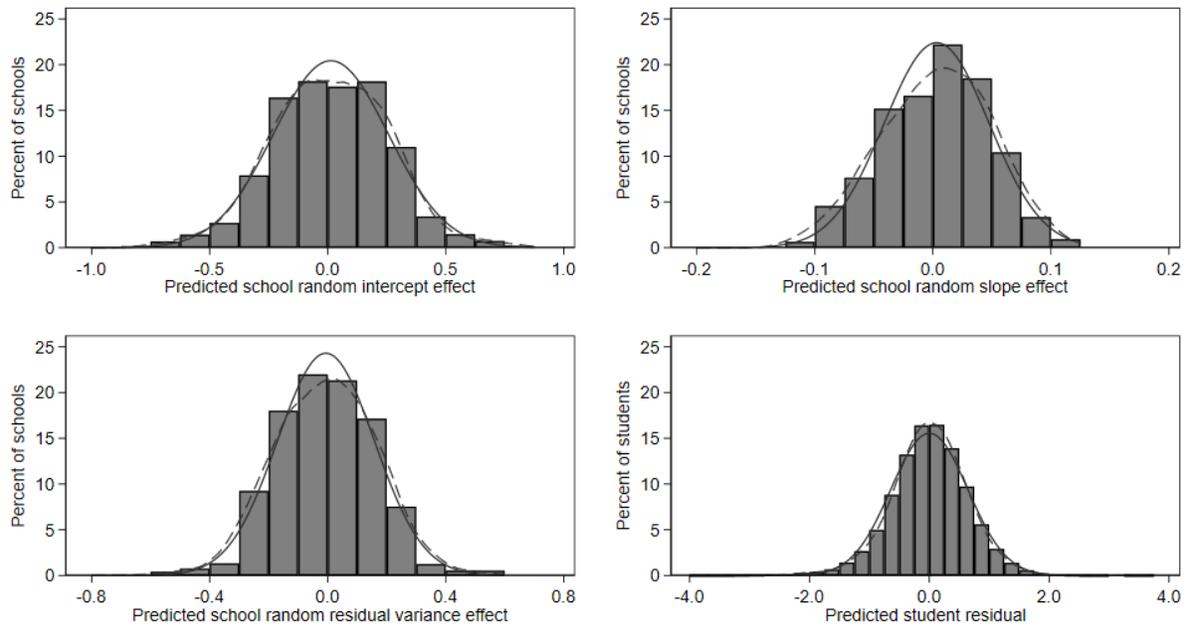

**FIGURE S5.** *Model 5 histograms of predicted school random intercept effects (top left), school random slope effects (top right), school residual variance effects (bottom left), and student residuals (bottom right), each with superimposed normal curves (solid) and kernel density curves (dashed).*

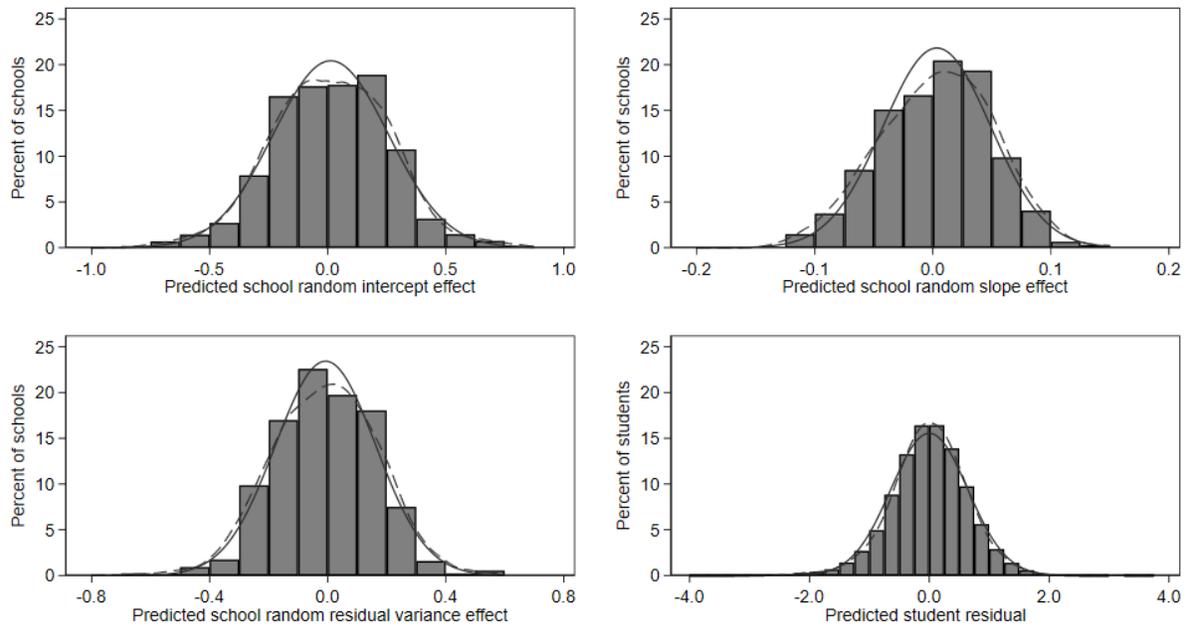

**FIGURE S6.** *Model 6 histograms of predicted school random intercept effects (top left), school random slope effects (top right), school residual variance effects (bottom left), and student residuals (bottom right), each with superimposed normal curves (solid) and kernel density curves (dashed).*

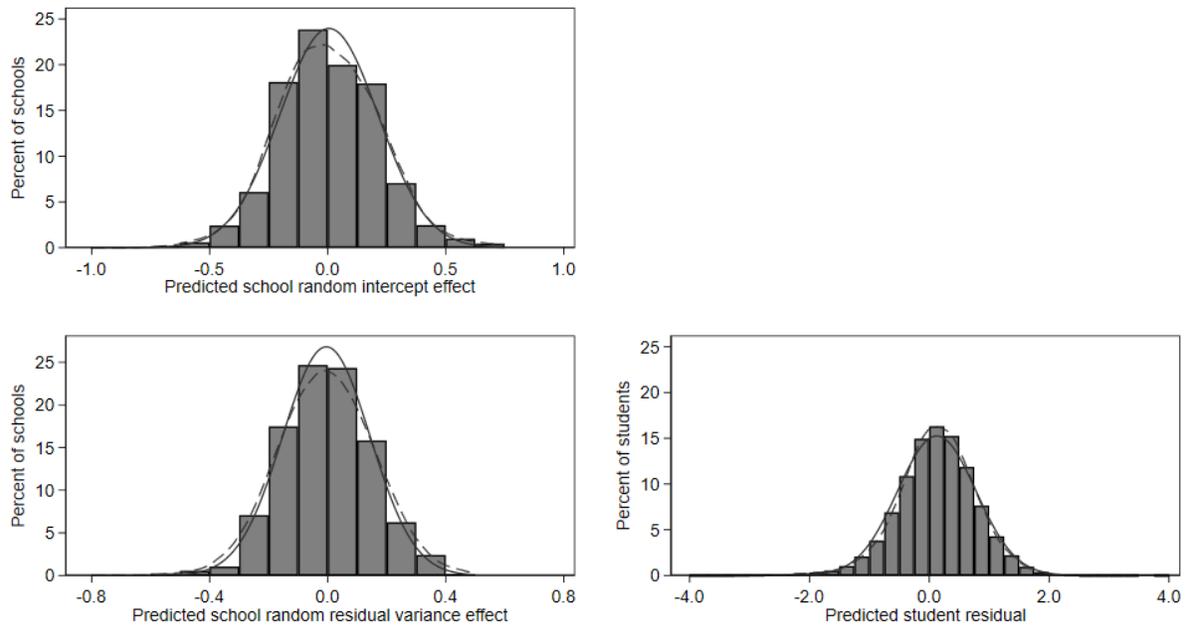

**FIGURE S7.** *Model 7 histograms of predicted school random intercept effects (top left), school random residual variance effects (bottom left), and student residuals (bottom right), each with superimposed normal curves (solid) and kernel density curves (dashed).*

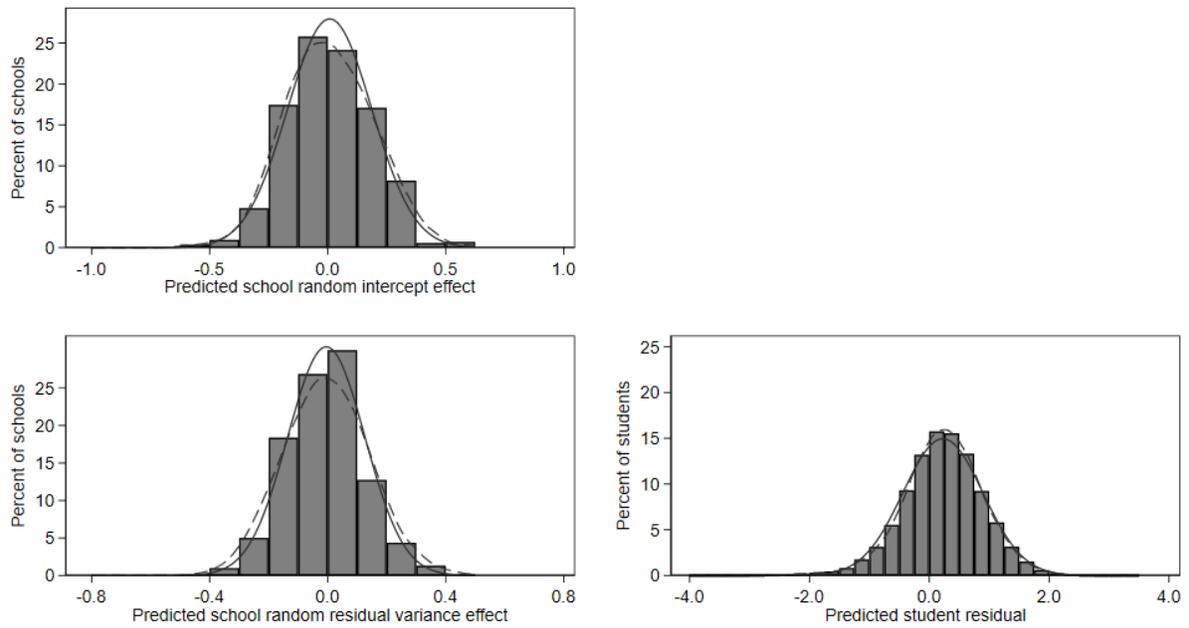

**FIGURE S8.** *Model 8 histograms of predicted school random intercept effects (top left), school random residual variance effects (bottom left), and student residuals (bottom right), each with superimposed normal curves (solid) and kernel density curves (dashed).*

## S2. Stata, R and MixWILD Software Instructions and Simulated Data for Fitting the Models

In this section, we describe Stata and R syntax to fit the models explored in this article by MCMC methods and MixWILD point-and-click instructions to fit these models by maximum likelihood estimation. To support readers, we provide script files and data to replicate the presented analysis.

### Example Model

For simplicity, we focus on the two-level random-intercept model with a random residual variance function presented in Section 3. To illustrate the syntax as simply as possible, we consider a version of this model with only one student characteristic (student prior achievement). This model can be written as

$$y_{ij} = \beta_0 + \beta_1 x_{ij} + u_j + e_{ij} \tag{S1}$$

$$\ln(\sigma_{e,ij}^2) = \alpha_0 + \alpha_1 x_{ij} + v_j$$

$$\begin{pmatrix} u_j \\ v_j \end{pmatrix} \sim N \left\{ \begin{pmatrix} 0 \\ 0 \end{pmatrix}, \begin{pmatrix} \sigma_u^2 & \\ \sigma_{uv} & \sigma_v^2 \end{pmatrix} \right\}$$

$$e_{ij} \sim N(0, \sigma_{e,ij}^2)$$

This model is the same as Model 3 presented in the article. Model 1 presented in the article can be viewed as a constrained version of this model (where the residual variance is assumed constant across all students and schools). Models 2, 5, and 6 vary in the covariates included in both the mean and residual variance functions. Model 4 is a random-slope version of this model (where a random slope is added to prior achievement).

**Simulated Data**

As we cannot share the data analyzed in the article, we analyze here simulated data where we use the above model as the data generating model. We simulate a single dataset with 100 schools and 25 students per school. We simulate $x_{ij}$ as standard normal variate with intraclass correlation of 0.2. We specify the true parameter values as $\beta_0 = 0$, $\beta_1 = 0.7$, $\sigma_u^2 = 0.05$, $\alpha_0 = -0.8$, $\alpha_1 = 0.05$, $\sigma_v^2 = 0.05$, $\sigma_{uv} = 0.025$. The resulting data can be found in the files data.dta and data.csv.

**Stata: The bayesmh Command**

We focus on the bayesmh Stata command (StataCorp, 2021). The bayesmh command implements an adaptive Metropolis-Hastings MCMC algorithm. We present the simplest possible syntax noting that mixing can be improved via model reparameterization (e.g., hierarchical centering) and by specifying various estimation options (initial values, blocking) and we encourage readers to consult the comprehensive documentation for further details.

The syntax to specify and fit this model is as follows.

```
. bayesmh y x U[school], ///
    likelihood(normal(exp({lnsigma2e:x,xb} + {V[school]}))) ///
    prior({y:}, normal(0, 10000)) ///
    prior({lnsigma2e:}, normal(0, 10000)) ///
    prior({U} {V}, mvnormal(2, 0, 0, {SIGMAUV, matrix})) ///
    prior({SIGMAUV, matrix}, iwishart(2, 3, S))
```

Line 1 of the syntax specifies the mean function. Line 2 specifies the normal response distribution and the residual variance function. The intercept is included in both functions by default. Lines 3 and 4 specify diffuse normal priors for the regression coefficients in each function with means of 0 and variances of 10000. Line 5 specifies the random effects to be bivariate normally distributed with zero means and a constant covariance matrix. Line 6 specifies a minimally informative inverse Wishart distribution for this covariance matrix (where S is pre-specified matrix such as an identity matrix).

The associated model output is as follows

```
Burn-in 2500 aaaaaaaaa1000aaaaaaaaa2000aaaaa done
Simulation 10000
.........1000.........2000.........3000.........4000.........5000.........6000......
> ...7000.........8000.........9000.........10000 done

Model summary
------------------------------------------------------------------------------
Likelihood:
  y ~ normal(xb_y,exp({lnsigma2e:x,xb}) + {V[school]}))

Prior:
  {y:x _cons} ~ normal(0,10000)                                              (1)

Hyperpriors:
    {lnsigma2e:x _cons} ~ normal(0,10000)
  {U[school] V[school]} ~ mvnormal(2,0,0,{SIGMAUV,m})
          {SIGMAUV,m} ~ iwishart(2,3,S)
------------------------------------------------------------------------------
(1) Parameters are elements of the linear form xb_y.

Bayesian normal regression                      MCMC iterations  =     12,500
Random-walk Metropolis-Hastings sampling        Burn-in          =      2,500
                                                MCMC sample size =     10,000
                                                Number of obs    =      2,500
                                                Acceptance rate  =       .196
                                                Efficiency:  min =    .002048
                                                             avg =     .01894
Log marginal-likelihood                                      max =     .04904

------------------------------------------------------------------------------
             |                                            Equal-tailed
             |      Mean   Std. dev.     MCSE     Median  [95% cred. interval]
-------------+----------------------------------------------------------------
y            |
           x |  .7005516   .0142504   .000858   .7005228   .6731723   .7301972
       _cons | -.0282751   .0220327   .001753  -.0289124  -.0706135   .0183987
```

```
-------------+----------------------------------------------------------------
lnsigma2e    |
           x |  .0520455   .0299735    .001694   .0520977  -.0059208   .1090707
        _cons | -.7796784   .0347738     .00157  -.7794015  -.8489381  -.7097778
-------------+----------------------------------------------------------------
 SIGMAUV_1_1 |  .0340151   .0068176    .001046   .0338658   .0200429   .0490111
 SIGMAUV_2_1 |  .0137237   .0077867    .001553   .0138888  -.0011668   .0284148
 SIGMAUV_2_2 |  .0431452   .0209756    .004636   .0395217   .0092352   .0902076
-------------+----------------------------------------------------------------
Note: There is a high autocorrelation after 500 lags.
Note: Adaptation tolerance is not met in at least one of the blocks.
```

The command ran one chain with 2500 burn-in iterations and 10000 monitoring iterations. The reader should note the warning messages at the end of the output. As noted above, mixing can be improved via model reparameterization and by specifying various estimation options. The results presented in tabular form are as follows

|  | True value | Est. | SE |
|---|---|---|---|
| $\beta_0$ | 0.000 | -0.028 | 0.022 |
| $\beta_1$ | 0.700 | 0.701 | 0.014 |
| $\sigma_u^2$ | 0.050 | 0.034 | 0.007 |
| $\alpha_0$ | -0.800 | -0.780 | 0.035 |
| $\alpha_1$ | 0.050 | 0.052 | 0.030 |
| $\sigma_v^2$ | 0.050 | 0.043 | 0.021 |
| $\sigma_{uv}$ | 0.025 | 0.014 | 0.008 |

The parameter estimates are similar to their true values and to those provided by brms in R and MixWILD (see below).

## R: The brms Package

We focus on the brm function of the brms R package (Bürkner, 2017, 2018). The brms package calls the Stan software (Stan Development Team, 2021) which implements Hamiltonian Monte Carlo (HMC) and no-U-turn samplers (NUTS). We present the simplest possible syntax noting that mixing can be improved via model reparameterization and by specifying various estimation options and we encourage readers to consult the comprehensive documentation for further details.

The brms package specifies model S1 using the following alternative parameterization.

$$y_{ij} = \beta_0 + \beta_1 x_{ij} + u_j + e_{ij} \tag{S2}$$

$$\ln(\sigma_{e,ij}) = \alpha'_0 + \alpha'_1 x_{ij} + v'_j$$

$$\begin{pmatrix} u_j \\ v'_j \end{pmatrix} \sim N\left\{ \begin{pmatrix} 0 \\ 0 \end{pmatrix}, \begin{pmatrix} \sigma_u^2 & \\ \sigma_{uv'} & \sigma_{v'}^2 \end{pmatrix} \right\}$$

Thus, the residual variance function is now specified in terms of the residual SD. Fortunately, the parameter and random effect values of the original parameterization can be easily recovered as follows $\alpha_0 = 2\alpha'_0$, $\alpha_1 = 2\alpha'_1$, $v_j = 2v'_j$, $\sigma_v^2 = 4\sigma_{v'}^2$.

The syntax to specify and fit this model is as follows

```
brm(bf(y ~ 1 + x + (1 |s| school),
       sigma ~ 1 + x + (1 |s| school)),
    data = mydata,
    family = gaussian()
)
```

where for further simplicity we use the default priors for all model parameters and random effects. These include improper flat priors for the regression coefficients, half student-t priors with three degrees of freedom for the random effect standard deviations, and the LKJcorr prior for random effect correlation matrix (Bürkner, 2017).

Line 1 of the brm function syntax specifies the mean function. Line 2 specifies the

residual variance function parameterized in terms of the residual SD rather than the residual

variance. In these two lines, the "|s|" is used to indicate that the mean function and residual

variance function random effects are part of the same set and will therefore by default be allowed to correlate. Line 3 specifies the data frame. Line 4 specifies the normal response distribution.

The associated model output is as follows.

```
Compiling Stan program...
Start sampling

SAMPLING FOR MODEL '93a90408567ae7343eea598de7d7e540' NOW (CHAIN 1).
Chain 1:
Chain 1: Gradient evaluation took 0.005 seconds
Chain 1: 1000 transitions using 10 leapfrog steps per transition would take 50 seconds.
Chain 1: Adjust your expectations accordingly!
Chain 1:
Chain 1:
Chain 1: Iteration:    1 / 2000 [  0%]  (Warmup)
Chain 1: Iteration:  200 / 2000 [ 10%]  (Warmup)
Chain 1: Iteration:  400 / 2000 [ 20%]  (Warmup)
Chain 1: Iteration:  600 / 2000 [ 30%]  (Warmup)
Chain 1: Iteration:  800 / 2000 [ 40%]  (Warmup)
Chain 1: Iteration: 1000 / 2000 [ 50%]  (Warmup)
Chain 1: Iteration: 1001 / 2000 [ 50%]  (Sampling)
Chain 1: Iteration: 1200 / 2000 [ 60%]  (Sampling)
Chain 1: Iteration: 1400 / 2000 [ 70%]  (Sampling)
Chain 1: Iteration: 1600 / 2000 [ 80%]  (Sampling)
Chain 1: Iteration: 1800 / 2000 [ 90%]  (Sampling)
Chain 1: Iteration: 2000 / 2000 [100%]  (Sampling)
Chain 1:
Chain 1:  Elapsed Time: 41.769 seconds (Warm-up)
Chain 1:                19.914 seconds (Sampling)
Chain 1:                61.683 seconds (Total)
Chain 1:

SAMPLING FOR MODEL '93a90408567ae7343eea598de7d7e540' NOW (CHAIN 2).
Chain 2:
Chain 2: Gradient evaluation took 0.001 seconds
Chain 2: 1000 transitions using 10 leapfrog steps per transition would take 10 seconds.
Chain 2: Adjust your expectations accordingly!
Chain 2:
Chain 2:
Chain 2: Iteration:    1 / 2000 [  0%]  (Warmup)
Chain 2: Iteration:  200 / 2000 [ 10%]  (Warmup)
Chain 2: Iteration:  400 / 2000 [ 20%]  (Warmup)
Chain 2: Iteration:  600 / 2000 [ 30%]  (Warmup)
Chain 2: Iteration:  800 / 2000 [ 40%]  (Warmup)
Chain 2: Iteration: 1000 / 2000 [ 50%]  (Warmup)
Chain 2: Iteration: 1001 / 2000 [ 50%]  (Sampling)
Chain 2: Iteration: 1200 / 2000 [ 60%]  (Sampling)
Chain 2: Iteration: 1400 / 2000 [ 70%]  (Sampling)
```

```
Chain 2: Iteration: 1600 / 2000 [ 80%]  (Sampling)
Chain 2: Iteration: 1800 / 2000 [ 90%]  (Sampling)
Chain 2: Iteration: 2000 / 2000 [100%]  (Sampling)
Chain 2:
Chain 2:  Elapsed Time: 49.423 seconds (Warm-up)
Chain 2:                21.946 seconds (Sampling)
Chain 2:                71.369 seconds (Total)
Chain 2:

SAMPLING FOR MODEL '93a90408567ae7343eea598de7d7e540' NOW (CHAIN 3).
Chain 3:
Chain 3: Gradient evaluation took 0 seconds
Chain 3: 1000 transitions using 10 leapfrog steps per transition would take 0 seconds.
Chain 3: Adjust your expectations accordingly!
Chain 3:
Chain 3:
Chain 3: Iteration:    1 / 2000 [  0%]  (Warmup)
Chain 3: Iteration:  200 / 2000 [ 10%]  (Warmup)
Chain 3: Iteration:  400 / 2000 [ 20%]  (Warmup)
Chain 3: Iteration:  600 / 2000 [ 30%]  (Warmup)
Chain 3: Iteration:  800 / 2000 [ 40%]  (Warmup)
Chain 3: Iteration: 1000 / 2000 [ 50%]  (Warmup)
Chain 3: Iteration: 1001 / 2000 [ 50%]  (Sampling)
Chain 3: Iteration: 1200 / 2000 [ 60%]  (Sampling)
Chain 3: Iteration: 1400 / 2000 [ 70%]  (Sampling)
Chain 3: Iteration: 1600 / 2000 [ 80%]  (Sampling)
Chain 3: Iteration: 1800 / 2000 [ 90%]  (Sampling)
Chain 3: Iteration: 2000 / 2000 [100%]  (Sampling)
Chain 3:
Chain 3:  Elapsed Time: 44.674 seconds (Warm-up)
Chain 3:                18.936 seconds (Sampling)
Chain 3:                63.61 seconds (Total)
Chain 3:

SAMPLING FOR MODEL '93a90408567ae7343eea598de7d7e540' NOW (CHAIN 4).
Chain 4:
Chain 4: Gradient evaluation took 0.001 seconds
Chain 4: 1000 transitions using 10 leapfrog steps per transition would take 10 seconds.
Chain 4: Adjust your expectations accordingly!
Chain 4:
Chain 4:
Chain 4: Iteration:    1 / 2000 [  0%]  (Warmup)
Chain 4: Iteration:  200 / 2000 [ 10%]  (Warmup)
Chain 4: Iteration:  400 / 2000 [ 20%]  (Warmup)
Chain 4: Iteration:  600 / 2000 [ 30%]  (Warmup)
Chain 4: Iteration:  800 / 2000 [ 40%]  (Warmup)
Chain 4: Iteration: 1000 / 2000 [ 50%]  (Warmup)
Chain 4: Iteration: 1001 / 2000 [ 50%]  (Sampling)
Chain 4: Iteration: 1200 / 2000 [ 60%]  (Sampling)
Chain 4: Iteration: 1400 / 2000 [ 70%]  (Sampling)
Chain 4: Iteration: 1600 / 2000 [ 80%]  (Sampling)
Chain 4: Iteration: 1800 / 2000 [ 90%]  (Sampling)
Chain 4: Iteration: 2000 / 2000 [100%]  (Sampling)
Chain 4:
Chain 4:  Elapsed Time: 54.744 seconds (Warm-up)
Chain 4:                20.939 seconds (Sampling)
Chain 4:                75.683 seconds (Total)
Chain 4:
 Family: gaussian
  Links: mu = identity; sigma = log
Formula: y ~ 1 + x + (1 | s | school)
         sigma ~ 1 + x + (1 | s | school)
   Data: mydata (Number of observations: 2500)
  Draws: 4 chains, each with iter = 2000; warmup = 1000; thin = 1;
         total post-warmup draws = 4000

Group-Level Effects:
~school (Number of levels: 100)
```

| | Estimate | Est.Error | l-95% CI | u-95% CI | Rhat | Bulk_ESS | Tail_ESS |
|---|---|---|---|---|---|---|---|
| sd(Intercept) | 0.19 | 0.02 | 0.15 | 0.24 | 1.00 | 1552 | 2474 |
| sd(sigma_Intercept) | 0.11 | 0.02 | 0.06 | 0.16 | 1.01 | 1333 | 1396 |

```
cor(Intercept,sigma_Intercept)     0.32      0.18     -0.04      0.67 1.00      2180      2381

Population-Level Effects:
                Estimate Est.Error l-95% CI u-95% CI Rhat Bulk_ESS Tail_ESS
Intercept          -0.03      0.02    -0.07     0.02 1.00     2188     2671
sigma_Intercept    -0.39      0.02    -0.43    -0.36 1.00     3600     3190
x                   0.70      0.01     0.67     0.73 1.00     6686     3234
sigma_x             0.03      0.02    -0.00     0.05 1.00     5863     3170

Draws were sampled using sampling(NUTS). For each parameter, Bulk_ESS
and Tail_ESS are effective sample size measures, and Rhat is the potential
scale reduction factor on split chains (at convergence, Rhat = 1).
```

The model ran four chains each with 1000 warmup (burn-in) iterations and 1000 monitoring iterations. Recall that the model is a reparametrized version of model S1. The outputs additionally shows that the elements of the random effect covariance matrix used in this alternative parameterizations are presented as SDs and correlations rather than as variances and covariances. We can recover the random effect variances of this alternative parameterization by squaring the random effect SDs. The random effect covariance of this alternative parameterization can be recovered by multiplying the random effect correlation by the two random effect SDs. We can then recover the parameter and random effect values associated with the parameterization S1 using the transformations listed previously. All these calculations are best applied to the underlying chains rather than the means which are displayed in the output. Having carried out these steps, the results are as follows.

|  | True value | Est. | SE |
|---|---|---|---|
| $\beta_0$ | 0.000 | -0.028 | 0.024 |
| $\beta_1$ | 0.700 | 0.701 | 0.014 |
| $\sigma_u^2$ | 0.050 | 0.038 | 0.008 |
| $\alpha_0$ | -0.800 | -0.784 | 0.037 |
| $\alpha_1$ | 0.050 | 0.050 | 0.030 |
| $\sigma_v^2$ | 0.050 | 0.054 | 0.022 |
| $\sigma_{uv}$ | 0.025 | 0.014 | 0.009 |

The parameter estimates are similar to their true values and to those provided by bayesmh in Stata (above) and MixWILD (below).

**MixWILD**

The MixWILD software (Dzubur et al., 2020) is freely available at the software website https://reach-lab.github.io/MixWildGUI/. MixWILD fits models using maximum likelihood estimation via adaptive quadrature. We use the default estimation options which specifies 11 quadrature points. We encourage readers to consult the comprehensive documentation for further details.

The MixWILD software specifies model S1 using the following alternative parameterization.

$$y_{ij} = \beta_0 + \beta_1 x_{ij} + u_j + e_{ij} \qquad \text{(S3)}$$

$$\ln(\sigma_u^2) = \gamma_0$$

$$\ln(\sigma_{e,ij}^2) = \alpha_0 + \alpha_1 x_{ij} + \tau \frac{u_j}{\sigma_u} + v_j'$$

$$\begin{pmatrix} u_j \\ v_j' \end{pmatrix} \sim N \left\{ \begin{pmatrix} 0 \\ 0 \end{pmatrix}, \begin{pmatrix} \sigma_u^2 & \\ 0 & \sigma_{v'}^2 \end{pmatrix} \right\}$$

The mean function random effect variance is specified on the log scale $\gamma_0$. The mean and residual function random effects are assumed independent. The association between the mean and residual variance functions is instead allowed for via entering the mean function random effect as a standardized latent covariate in the residual variance function and estimating its regression coefficient $\tau$. Fortunately, the parameter and random effect values of the original parameterization are easily recovered as follows $\sigma_u^2 = \exp(\gamma_0)$, $v_j = \frac{u_j}{\sigma_u} + v_j'$, $\sigma_v^2 = \tau^2 + \sigma_{v'}^2$,

$\sigma_{uv} = \tau \sigma_u$.

The point-and-click instructions to specify and fit this model are as follows.

- Open **MixWILD** by double clicking on the MixWILD icon

MixWILD will open and you will see the following **Model Configuration** tab

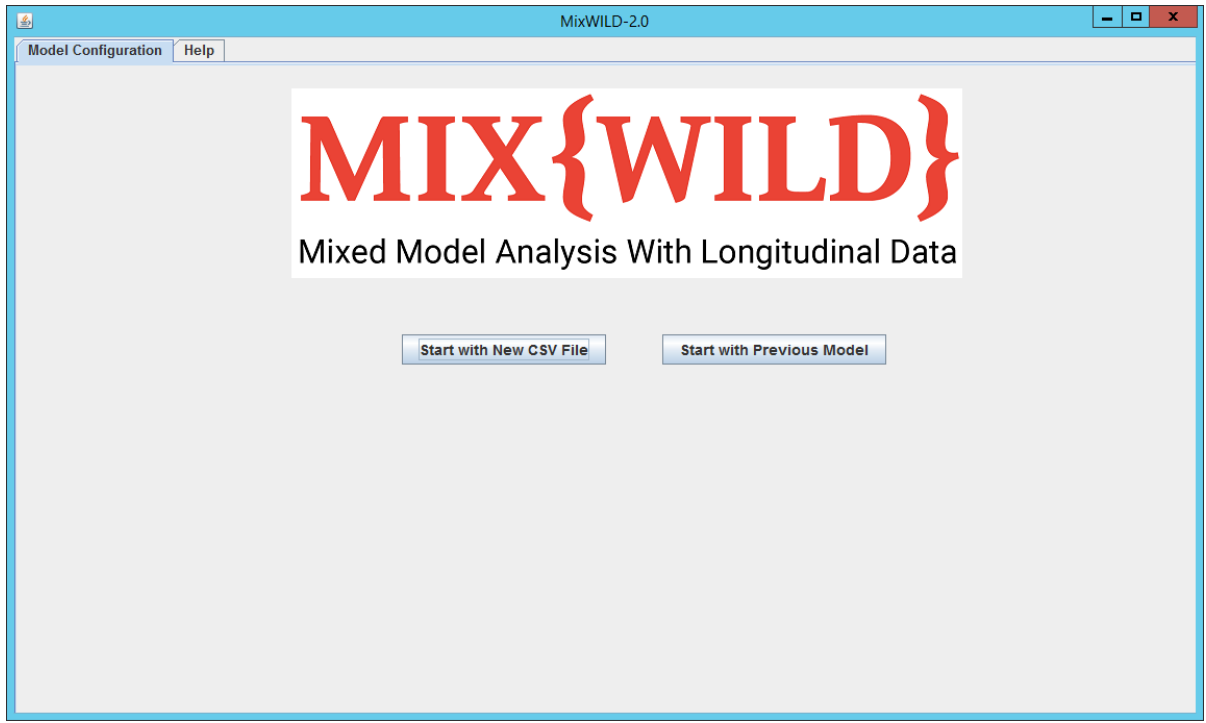

- Click **Start with New CSV File**

An **Open** file window will appear

- Navigate to wherever you have saved **data.csv** and click **Open** (please make sure there are no spaces in the file address!)

The **Model Configuration** tab will update to match that shown below

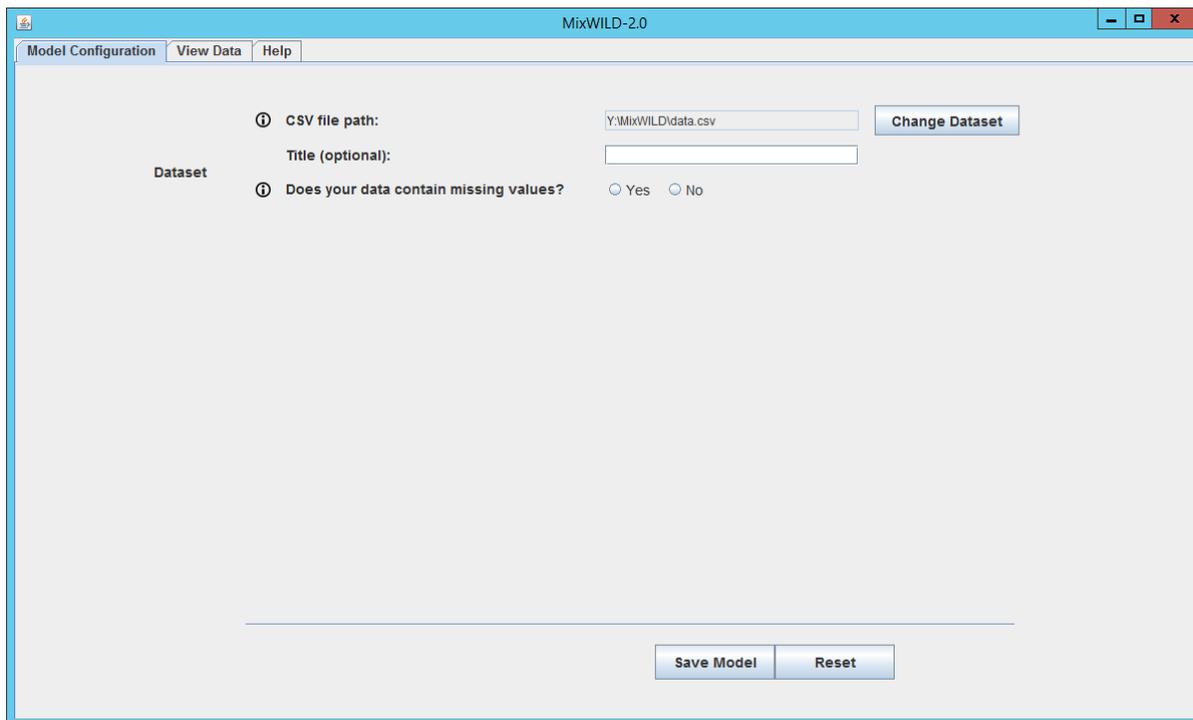

- Next to **Does your data contain missing values?**, select **No**

The **Model Configuration** tab will update to reveal more options as shown below.

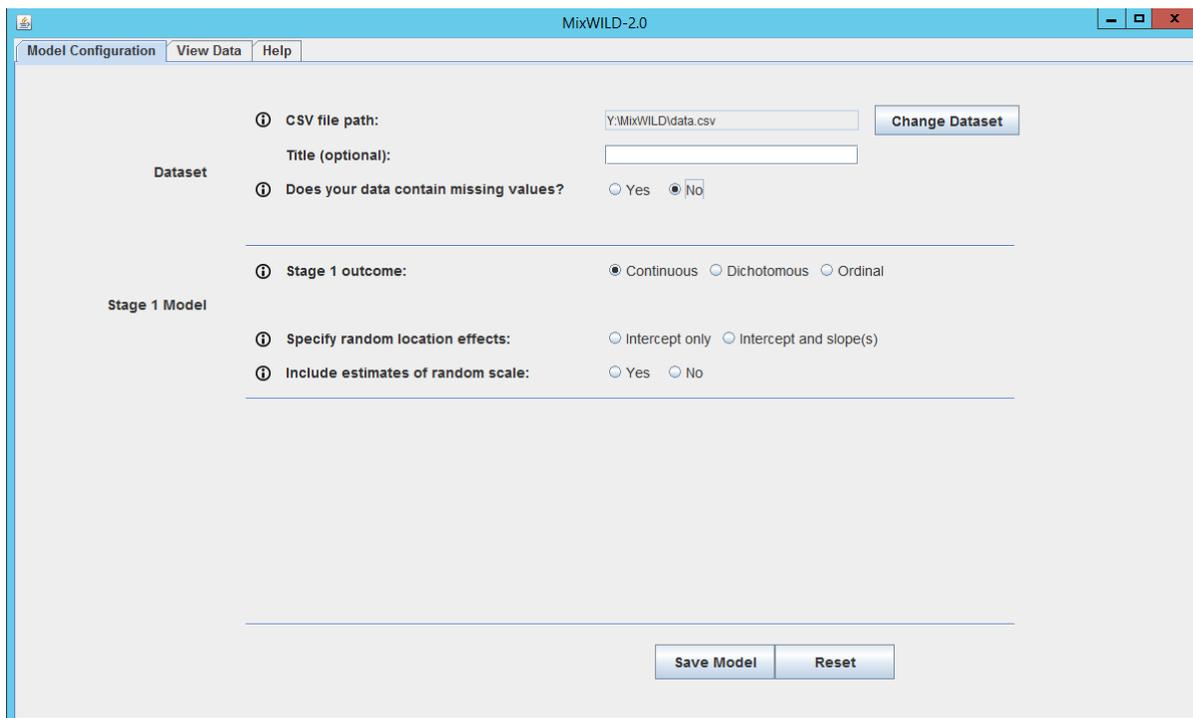

- Next to **Specify random location effects**, select **Intercept only**

- Next to **Include estimates of random scale**, select **Yes**

The **Model Configuration** tab will update to reveal more options as shown below.

- Next to **Include Stage 2 model**, select **No**

- Check that the **Model Configuration** tab matches that shown below, then click **Continue**

The **Stage 1 Configuration** tab will then be displayed

- Under the **Stage 1 Outcome** drop-down list, select **y**

- Click **Configure State 1 Regressors …**

The **Add Stage 1 Regressors** window will open

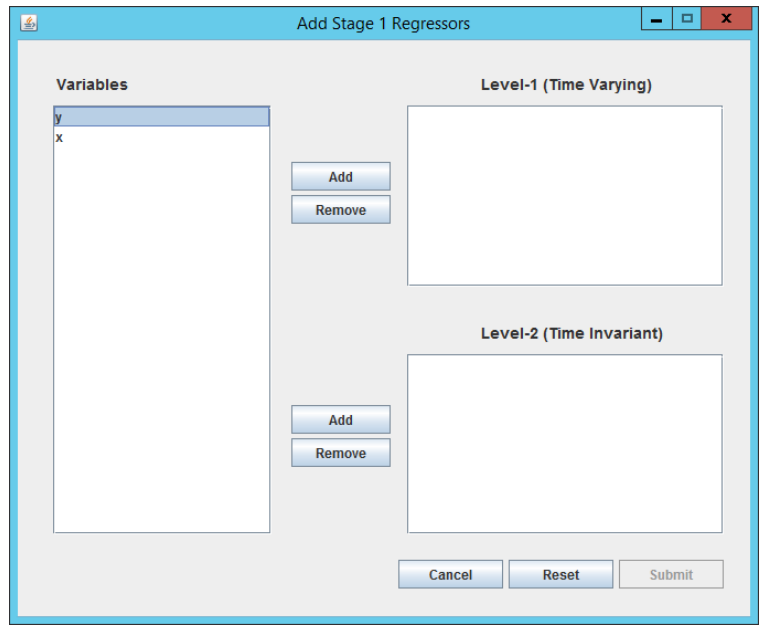

- In the **Variables** box, select **x**

- Next to the **Level-1 (Time Varying)** box, click **Add**

- Check that the **Add Stage 1 Regressors** tab matches that shown below, then Click

  **Submit**

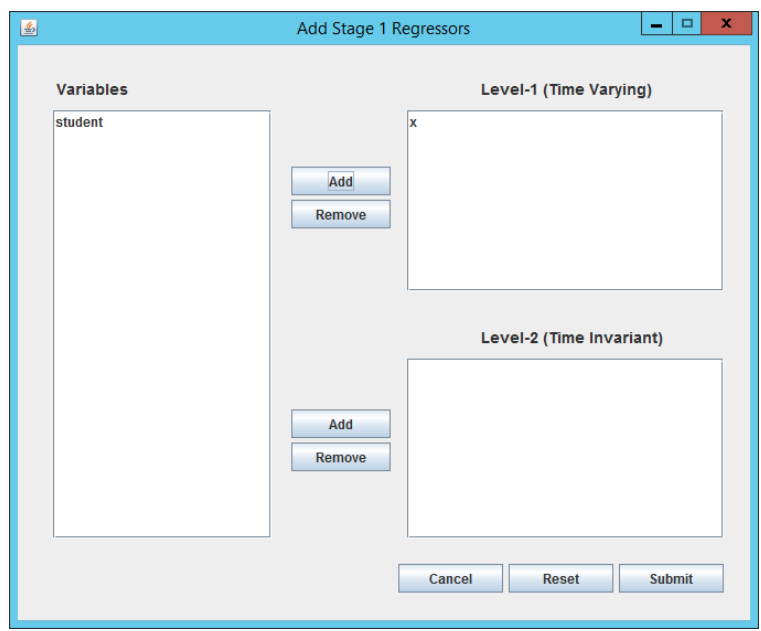

You will once again see the **Stage 1 Configuration** tab

- Under **Stage 1 Regressors**, under **Level-1**, check the two checkboxes with row label **x** and column labels **Mean** and **WS Variance**

- Under **Specify the relationship between the mean and WS variance**, select **Linear Association**

- Check the **Stage 1 Configuration** tab matches that shown below, then click **Run Stage 1**

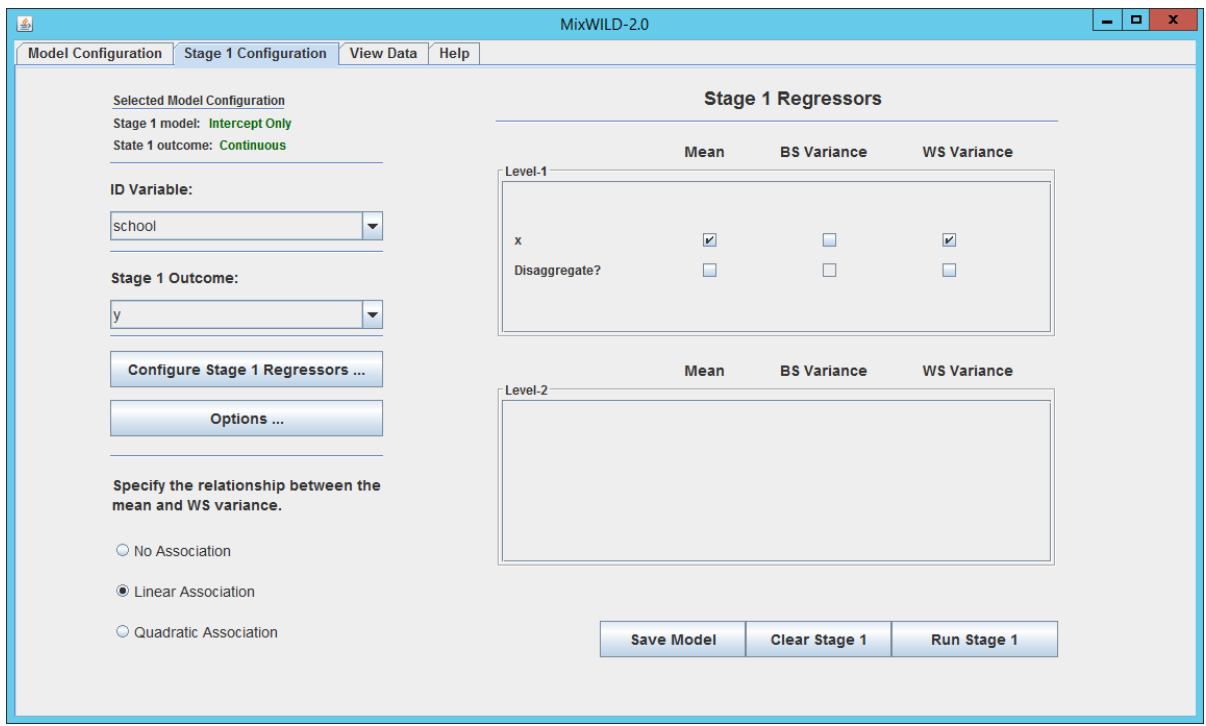

The following **Definition File Preview** window will open

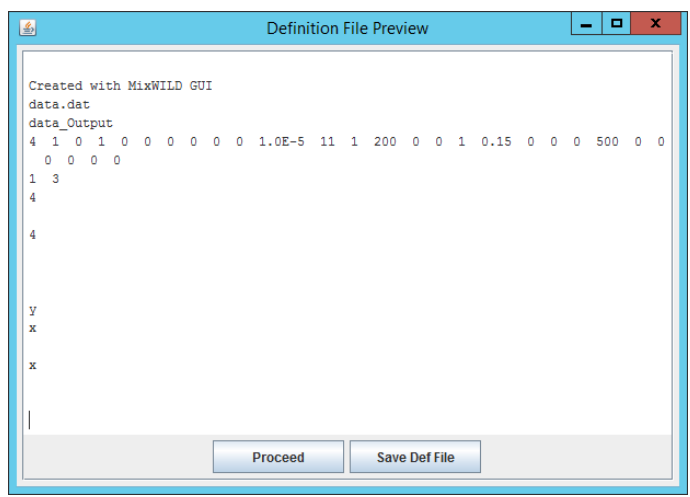

- Click **Proceed**

The following **Please wait … window** will appear

Once the estimation has completed, the window will automatically close and you will see the

following **Stage 1 Results** tab

The results presented in this window are pasted and discussed below

```
MIXREGLS_both: Mixed-effects Location Scale Model

-----------------------------
mixREGLS_both.DEF specifications
-----------------------------

 Created with MixWILD GUI

 data and output files:
 data.dat
 data_Output_stage1.out

 MULTIPLE LOCATION EFFECTS  = F
 SCALE EFFECT  = T
 CONVERGENCE CRITERION =  0.00001000
 RIDGEIN  =   0.1500
 NQ       =    11
 QUADRATURE =    1 (0=non-adaptive, 1=adaptive)
 MAXIT    =   200

------------
Descriptives
------------

 Number of level-1 observations =     2500

 Number of level-2 clusters     =      100

 Number of level-1 observations for each level-2 cluster
    25    25    25    25    25    25    25    25    25    25    25    25    25
    25    25    25    25    25    25    25    25    25    25    25    25    25
    25    25    25    25    25    25    25    25    25    25    25    25    25
    25    25    25    25    25    25    25    25    25    25    25    25    25
    25    25    25    25    25    25    25    25    25    25    25    25    25
    25    25    25    25    25    25    25    25    25    25    25    25    25
    25    25    25    25    25    25    25    25    25    25    25    25    25
    25    25    25    25    25    25    25    25    25
```

Dependent variable

| | mean | min | max | std dev |
|---|---|---|---|---|
| y | | -0.0366 | -3.4413 | 3.6288 | 1.0033 |

Mean model covariates

| | mean | min | max | std dev |
|---|---|---|---|---|
| intercept | 1.0000 | 1.0000 | 1.0000 | 0.0000 |
| x | -0.0124 | -4.5322 | 3.4890 | 1.0074 |

BS variance model covariates

| | mean | min | max | std dev |
|---|---|---|---|---|
| intercept | 1.0000 | 1.0000 | 1.0000 | 0.0000 |

WS variance model covariates

| | mean | min | max | std dev |
|---|---|---|---|---|
| intercept | 1.0000 | 1.0000 | 1.0000 | 0.0000 |
| x | -0.0124 | -4.5322 | 3.4890 | 1.0074 |

```
------------------------------
Model without Scale Parameters
------------------------------
 Total  Iterations =   4
 Final Ridge value = 0.2

 Log Likelihood                  =    -2649.111
 Akaike's Information Criterion =    -2653.111
 Schwarz's Bayesian Criterion   =    -2658.321

 ==> multiplied by -2
 Log Likelihood                  =     5298.221
 Akaike's Information Criterion =     5306.221
 Schwarz's Bayesian Criterion   =     5316.642
```

| Variable | Estimate | AsymStdError | z-value | p-value |
| --- | --- | --- | --- | --- |
| BETA (regression coefficients) | | | | |
| intercept | -0.02799 | 0.02358 | -1.18736 | 0.23509 |
| x | 0.70113 | 0.01453 | 48.24144 | 0.00000 |
| ALPHA (BS variance parameters: log-linear model) | | | | |
| intercept | -3.29934 | 0.21377 | -15.43434 | 0.00000 |
| TAU (WS variance parameters: log-linear model) | | | | |
| intercept | -0.76221 | 0.02887 | -26.40572 | 0.00000 |

```
---------------------------
Model WITH Scale Parameters
---------------------------
 Total  Iterations =  13
 Final Ridge value = 0.0

 Log Likelihood                  =    -2646.541
 Akaike's Information Criterion =    -2651.541
 Schwarz's Bayesian Criterion   =    -2658.054

 ==> multiplied by -2
 Log Likelihood                  =     5293.081
 Akaike's Information Criterion =     5303.081
 Schwarz's Bayesian Criterion   =     5316.107
```

| Variable | Estimate | AsymStdError | z-value | p-value |
| --- | --- | --- | --- | --- |
| BETA (regression coefficients) | | | | |
| intercept | -0.02778 | 0.02344 | -1.18525 | 0.23592 |
| x | 0.70271 | 0.01448 | 48.51461 | 0.00000 |
| ALPHA (BS variance parameters: log-linear model) | | | | |
| intercept | -3.31776 | 0.21505 | -15.42792 | 0.00000 |
| TAU (WS variance parameters: log-linear model) | | | | |
| intercept | -0.76310 | 0.02887 | -26.42848 | 0.00000 |
| x | 0.06373 | 0.02812 | 2.26655 | 0.02342 |

```
-----------------------
Model WITH RANDOM Scale
-----------------------
 Total  Iterations =  13
 Final Ridge value = 0.0
```

```
Log Likelihood                  =     -2639.033
Akaike's Information Criterion  =     -2646.033
Schwarz's Bayesian Criterion    =     -2655.151

==> multiplied by -2
Log Likelihood                  =      5278.066
Akaike's Information Criterion  =      5292.066
Schwarz's Bayesian Criterion    =      5310.302

Variable          Estimate      AsymStdError        z-value         p-value
----------------  ------------  ------------    ------------    ------------
BETA (regression coefficients)
intercept          -0.02783        0.02328        -1.19549         0.23190
x                   0.70159        0.01432        48.98862         0.00000
ALPHA (BS variance parameters: log-linear model)
intercept          -3.33908        0.21800       -15.31721         0.00000
TAU (WS variance parameters: log-linear model)
intercept          -0.78759        0.03724       -21.15057         0.00000
x                   0.05011        0.02971         1.68650         0.09170
Random scale standard deviation
Std Dev             0.21398        0.04480         4.77662         0.00000
Random location (mean) effect on WS variance
Loc  Eff            0.07740        0.04367         1.77238         0.07633

BS variance ratios and 95% CIs
------------------------------

Variable             Ratio          Lower           Upper
----------------  ------------  ------------    ------------
ALPHA (BS variance parameters: log-linear model)
intercept           0.03547        0.02314         0.05438

WS variance ratios and 95% CIs
------------------------------

Variable             Ratio          Lower           Upper
----------------  ------------  ------------    ------------
TAU (WS variance parameters: log-linear model)
intercept           0.45494        0.42292         0.48938
x                   1.05138        0.99191         1.11443
Random location (mean) effect on WS variance
Location Effect     1.08047        0.99184         1.17702

Random scale standard deviation
Std Dev             1.23859        1.13448         1.35226
```

The MixWILD output is presented in five sections. The first four sections present the estimation options, descriptive statistics, and the results of two simpler versions of the full model used to generating starting values for the full model. The fifth and final block of output titled "Model WITH RANDOM Scale" presents the results for the full model. Recall that the model is a reparametrized version of model S1. The output additionally shows that the residual variance function random effect variance used in this alternative parameterization is presented as a SD rather than a variance. We can recover the random effect variance of this alternative parameterization by squaring the random effect SD. We can then recover the parameter and random effect values associated with the original S1 parameterization using the transformations listed previously. The corresponding standard errors can be recovered via the delta method. Having carried out these steps, the results presented in tabular form are as follows

|  | True value | Est. | SE |
|---|---|---|---|
| $\beta_0$ | 0.000 | -0.028 | 0.023 |
| $\beta_1$ | 0.700 | 0.701 | 0.014 |
| $\sigma_u^2$ | 0.050 | 0.035 | 0.008 |
| $\alpha_0$ | -0.800 | -0.785 | 0.037 |
| $\alpha_1$ | 0.050 | 0.050 | 0.030 |
| $\sigma_v^2$ | 0.050 | 0.052 | 0.020 |
| $\sigma_{uv}$ | 0.025 | 0.015 | 0.008 |

The parameter estimates are similar to their true values and to those provide by bayesmh in Stata and brms in R (above).